\documentclass[12pt]{article}

\usepackage[backend=biber, style=apa]{biblatex}

\usepackage[margin=0.7in]{geometry}
\usepackage{xcolor,soul,framed} 

\colorlet{shadecolor}{yellow}
\usepackage[pdftex]{graphicx}

\usepackage[cmex10]{amsmath}
\usepackage{array}
\usepackage{mdwmath}
\usepackage{mdwtab}
\usepackage{eqparbox}
\usepackage{url}
\usepackage{textcomp}
\usepackage{amsmath,amssymb,amsfonts}
\usepackage{algorithmicx}
\usepackage{algorithm, algpseudocode}
\usepackage{hyperref}

\date{}

\usepackage{setspace}

\begin{document}

\clearpage

\title{\Large Motion-Invariant Variational Auto-Encoding of Brain Structural Connectomes}
\author{
Yizi Zhang\footnote{Yizi Zhang, Department of Statistics, Columbia University. Email: yz4123@columbia.edu.}, 
Meimei Liu\footnote{Meimei Liu, Department of Statistics, Virginia Polytechnic Institute and State University.}, 
Zhengwu Zhang\footnote{Zhengwu Zhang, Department of Statistics and Operations Research, The University of North Carolina at Chapel Hill.}, 
David Dunson\footnote{David Dunson, Department of Statistical Science, Duke University.}
}

\markboth{
}{}

\maketitle

\begin{doublespace} 
\begin{abstract}
Mapping of human brain structural connectomes via diffusion MRI offers a unique opportunity to understand brain structural connectivity and relate it to various human traits, such as cognition. However, \textcolor{black}{head displacement during image acquisition} can compromise the accuracy of connectome reconstructions and subsequent inference results. We develop a generative model to learn low-dimensional representations of structural connectomes invariant to \textcolor{black}{motion-induced artifacts}, so that we can link brain networks and human traits more accurately, and generate motion-adjusted connectomes. 
We apply the proposed model to data from the Adolescent Brain Cognitive Development (ABCD) study and the Human Connectome Project (HCP) to investigate how our motion-invariant connectomes facilitate understanding of the brain network and its relationship with cognition. Empirical results demonstrate that the proposed motion-invariant variational auto-encoder (inv-VAE) outperforms its competitors in various aspects. In particular, motion-adjusted structural connectomes are more strongly associated with a wide array of cognition-related traits than other approaches without motion adjustment. Code: {\textcolor{blue}{\texttt{https://github.com/yzhang511/inv-vae}.}}\footnote{This work is licensed under a Creative Commons Attribution 4.0 International License. For more details please see \href{https://creativecommons.org/licenses/by/4.0/}{https://creativecommons.org/licenses/by/4.0/}}.
\end{abstract}


\section{Introduction}
\label{sec:introduction}
To comprehensively understand brain function and organization, enormous efforts have been dedicated to imaging and analyzing the brain structural connectome, defined as a collection of white matter fiber tracts connecting different brain regions. Recent advancements in non-invasive neuroimaging techniques have led to a rapid increase in brain imaging datasets, such as the Human Connectome Project (HCP) \autocite{glasser2013} and the Adolescent Brain Cognitive Development (ABCD) study \autocite{bjork2017}. These developments have inspired a substantial body of literature \autocite{zhang2018a, zhang2019} that focuses on analyzing structural connectomes derived from diffusion-weighted magnetic resonance imaging (dMRI) data. However, these studies often encounter challenges due to subject head displacement during image acquisition, resulting in image misalignment and artifacts that hinder unbiased connectome reconstructions and analysis \autocite{andersson2016a}. Addressing these biases is crucial, which has motivated us to develop novel statistical methods capable of modeling and analyzing brain structural connectomes in an invariant manner \autocite{zemel2013, achille2018}.

\textcolor{black}{Our goal is to learn a low-dimensional representation of brain connectomes invariant to motion artifacts in the data, thereby facilitating prediction and inference on the relationship between brain structure and human traits such as cognitive abilities. A common motion correction technique in neuroimaging involves registering motion-corrupted images to a geometrically correct template \autocite{kochunov2006, bhushan2015}. However, in dMRI, motion frequently causes geometric distortion and signal dropout, significantly complicating registration \autocite{ben-amitay2012} and impacting subsequent brain network analyses \autocite{lebihan2006artifacts}. Current motion correction methods often utilize the FSL eddy tool \autocite{andersson2016b} to estimate and correct for eddy current-induced distortions and subject movements. Recent advancements include techniques like SHORELine \autocite{cieslak2021qsiprep} and SHARD \autocite{christiaens2021scattered} for correcting multi-shell dMRI sequences. Despite these improvements, studies have shown that even after motion correction in diffusion signals, in-scanner head motion can still have notable impacts on the recovered structural connectivity \autocite{baum2018impact}. Moreover, not all sources of image misalignment are due to motion \autocite{yendiki2014spurious}. Various imperfections in the imaging system can lead to artifacts causing image misalignment. Given these challenges, our major motivation is to develop a method that addresses motion-induced artifacts and adjusts for residual misalignment during the brain network modeling process, rather than in the data preprocessing stage. Here, motion-induced artifacts refer to the in-scanner head displacement that can be approximately quantified using image registration algorithms \autocite{andersson2016b}. Residual misalignment encompasses some remaining artifacts measured after eddy motion correction. By incorporating these considerations into our modeling approach, we aim to improve the robustness and reliability of brain connectome analysis in the presence of motion-induced artifacts.}

In biomedical imaging, removing unwanted effects in high dimensional data has been an important problem for at least the last decade. Without proper adjustment, the association between the signal of interest and one or more nuisance variables can degrade the ability to infer the signal accurately. This issue is particularly evident in brain imaging studies, and various tools have been developed. A popular matrix factorization method \autocite{alter2000} uses a singular value decomposition of the original matrix to filter out the singular vector associated with the nuisance variable, and reconstructs an adjusted matrix with the remaining singular vectors. Similarly, the Sparse Orthogonal to Group (SOG)  \autocite{aliverti2021} proposes an adjusted dataset based on a constrained form of matrix decomposition. In addition, model-based methods are popular alternatives for removing unwanted associations in the data. For example, distance-weighted discrimination \autocite{benito2004} adapts Support Vector Machines (SVM) for batch effect removal; the ComBat method \autocite{johnson2007} adjusts for batch effects in data using empirical Bayes. Our work falls under the model-based framework, and we are particularly interested in incorporating batch effect removal into a deep learning framework so that the non-linear batch effect can be modeled and removed. 

Removing unwanted information in the deep learning literature is formulated as an invariant representation learning problem. \textcolor{black}{Existing deep learning methods learn invariant representations through either adversarial or non-adversarial methods. Adversarial learning methods, prevalent in domain adaptation, involve a generative adversarial network (GAN) \autocite{goodfellow2020generative} where the generator extracts features to deceive the discriminator, while the discriminator aims to distinguish between different data domains. However, adversarial learning has limitations, including GANs' mode collapse issues and sensitivity to hyperparameter choices \autocite{papernot2016limitations}. Conversely, non-adversarial invariant methods, including recent group invariance techniques like contrastive learning \autocite{chen2020simple, he2020momentum, khosla2020supervised} and conditionally invariant VAE \autocite{aliee2023conditionally}, aim to develop predictors invariant to a group's action on the input space. Despite their strengths, these methods face a notable limitation: they are designed to handle categorical nuisance variables and may encounter challenges when confronted with continuous-valued nuisance variables. Some deep learning methods aim to mitigate the impact of continuous-valued nuisances, often incorporating mutual information loss. For instance, \autocite{greenfeld2020robust} utilize the Hilbert Schmidt Independence Criterion (HSIC), while \autocite{yu2021measuring} replace HSIC with matrix-based mutual information, yielding performance improvements. Another variant, squared-loss MI (SMI) \autocite{liu2021lsmi}, is also popular in this context. However, these mutual information-based methods focus solely on capturing dependencies in data in an unsupervised setting and don't ensure that the learned representations are task-relevant. In our work, we advocate for learning motion-invariant representations guided by the information bottleneck principle \autocite{alemi2016deep, moyer2018}. Our approach removes the mutual information between latent representations and \textcolor{black}{undesirable artifacts}, with latent representations learned via graph variational autoencoders (VAE) \autocite{liu2019}. Situated in the information bottleneck literature, our work addresses the challenge of finding representations that maximize the relation with the target response while minimizing mutual information with nuisance variables \textcolor{black}{like subject motion and residual image misalignment}. This framework not only addresses the noted limitations of existing methods but also correlates learned invariant representations with human behavior and cognition-related attributes.}

We aim to model brain connectomes and learn their representations invariant to nuisance factors, particularly motion-related artifacts. Our analysis of ABCD and HCP data, preprocessed with FSL eddy software \autocite{andersson2016b}, reveals that structural connectome data are significantly affected by motion. This impact is observed not only in the motion quantified by FSL eddy but also in residual misalignment in the diffusion MRI data after FSL eddy correction. Notably, individuals exhibiting substantial head displacement during data acquisition tend to have fewer reconstructed fiber connections across nearly all brain regions.
To address this issue, we develop a generative model that learns the distribution of brain connectomes while adjusting for undesirable artifacts induced by motion. This model also learns low-dimensional representations of structural connectomes that are invariant to motion. We achieve this by introducing a penalized VAE objective to eliminate mutual information between connectome representations and motion measures.
The learned invariant representation enhances our understanding of motion's impact on structural connectivity reconstructions. It can also be used to produce motion-adjusted connectomes for downstream analysis tasks, such as predicting behavior- and cognition-related traits and inferring structural connections between brain regions. Our analysis of ABCD and HCP data demonstrates that these motion-invariant representations improve the prediction of many cognitive traits by 5\% to 25\% compared to connectome representations without motion adjustment.

\textcolor{black}{The rest of the paper is organized as follows: Section \ref{sec:data} introduces the dMRI datasets used for deriving structural brain connectomes and outlines the motion measures utilized in this study. In Section \ref{sec:methods}, we present our method for adjusting for \textcolor{black}{motion-induced} artifacts during the modeling of structural brain networks. Section \ref{sec:simulation} details a simulation study verifying the effectiveness of our method in removing \textcolor{black}{unwanted information}. Section \ref{sec:applications} applies our proposed method to two large neuroimaging studies, demonstrating its efficacy in correcting for \textcolor{black}{motion-induced} artifacts and establishing a more accurate analysis between structural connectomes and cognitive traits. Section \ref{sec:discussion} includes a discussion. In the Appendix, we present additional results by considering the \textcolor{black}{residual misalignment after FSL eddy processing as an alternative source of artifacts}. Furthermore, we expand the simulation study to assess the robustness of our method when handling extreme \textcolor{black}{motion-related} artifacts in the Appendix.}

\section{Datasets Studied}\label{sec:data}

\subsection{Structural Connectomes and Cognitive Traits}
The ABCD study tracks the brain development of over 10,000 adolescents in the United States. 
We download and process the diffusion MRI (dMRI) data for 8,646 subjects from the 
NIH Data Archive. 
Details about data acquisition and preprocessing can be found in \autocite{casey2018}. 
The Human Connectome Project (HCP) characterizes brain connectivity in more than 1,000 adults; see \autocite{essen2012} for details about data acquisition.  
The latest dMRI data from the HCP can be accessed through ConnectomeDB. We download and process 1065 subjects from the HCP. 
\begin{figure}[t]
\centering
\includegraphics[width=.9\linewidth]{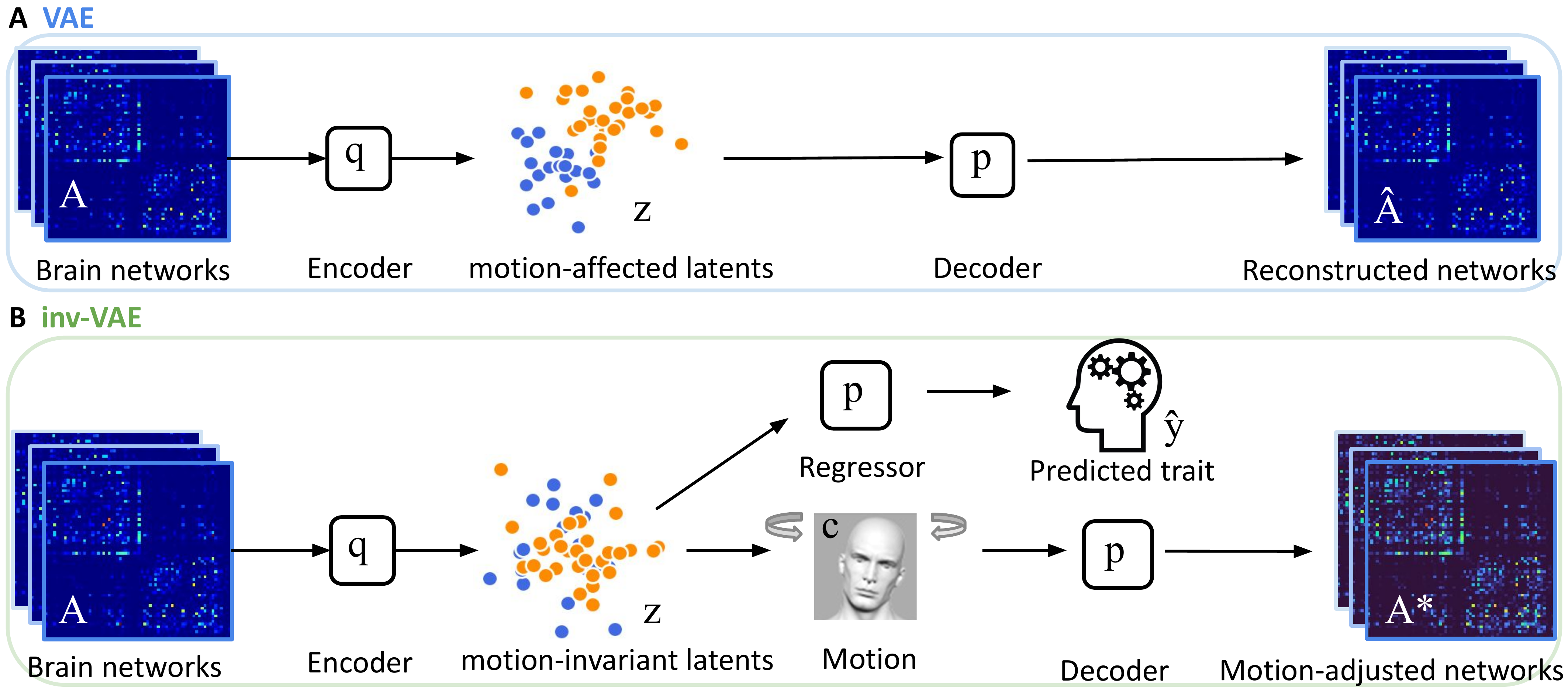}
\caption{
(\textbf{A}) Structure of the standard VAE model. The encoder learns motion-affected latents from motion-affected brain networks, and the decoder reconstructs networks from these latents. (\textbf{B}) Architecture of our proposed motion-invariant VAE. The encoder learns motion-invariant latents from the motion-affected networks, and the decoder reconstructs motion-adjusted networks from the invariant latents and the user-specified amount of subject motion. The invariant latents additionally serve as inputs to a regressor for predicting cognition-related traits.} \label{fig:model}
\end{figure}

To extract structural connectomes from the raw dMRI data, a reproducible tractography algorithm \autocite{maier2017} is used to generate the whole-brain tractography for each subject. We use the Desikan-Killiany parcellation \autocite{desikan2006} to define brain regions of interest (ROIs) corresponding to different nodes in the brain network. The Desikan-Killiany
atlas contains 68 ROIs with 34 ROIs belonging to each hemisphere. We use Freesurfer \autocite{iannopollo2019} to perform brain registration and parcellation. We extract streamlines connecting ROI pairs from the tractography data and the brain parcellation.  Fiber count is often used to measure the coupling strength between ROI pairs in the current literature \autocite{fornito2013, zhang2019}, and therefore we summarize each brain connection with its fiber count. 

ABCD and HCP use reliable measures to assess a wide range of human
behaviors and functions \autocite{gershon2013}. We focus on inferring the relationship between structural connectomes and cognition-related traits, e.g., scores from the picture vocabulary and oral reading recognition tests.

\begin{figure}[ht]
\centering
\includegraphics[width=.85\linewidth]{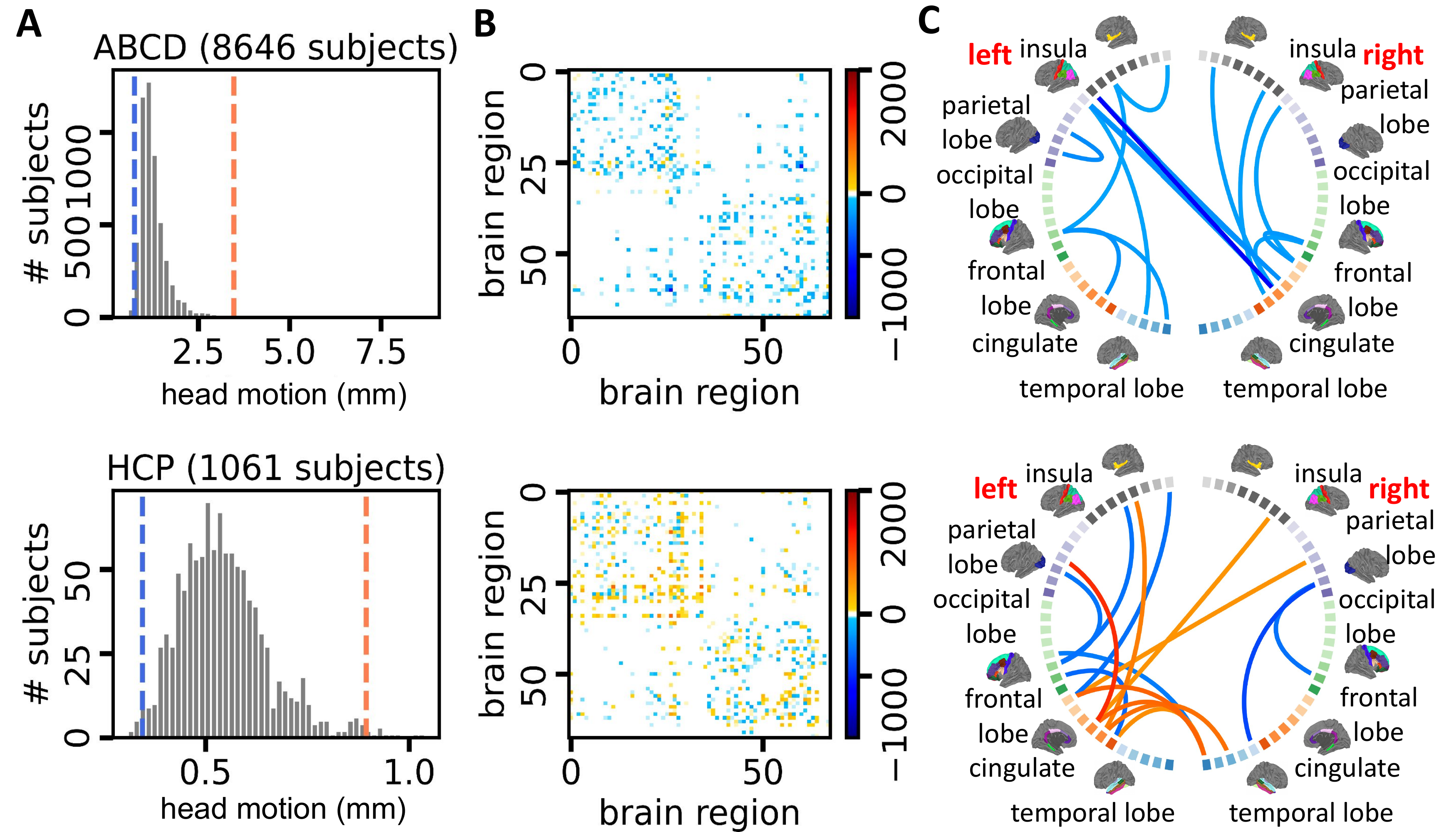}
\caption{Head motion analysis for subjects in the ABCD and HCP datasets and the impact of motion on brain structural connectome data. (\textbf{A}) Distribution histograms of head motion for ABCD and HCP subjects, with the blue and orange lines representing the 10th and 90th percentiles. Subjects with head motion below the blue line are classified into the small motion group, while those above the orange line belong to the large motion group. (\textbf{B}) Mean network differences between the large and small motion groups (large - small). (\textbf{C}) Circular plots representing the data in panel (B). The color scale corresponds to that in (B), with the display limited to the 15 edges showing the largest fiber count differences.}\label{fig:motion} 
\end{figure}

\subsection{Motion Quantification}\label{sec:motion_artifacts} 

\textcolor{black}{Our study considers two sources of motion artifacts: head motion and residual misalignment. Head motion refers to the displacement of the head in the scanner estimated by the eddy tool \autocite{andersson2016b}. Residual misalignment, on the other hand, refers to any remaining misalignment after applying the eddy correction tool. While we analyze both types of artifacts, our primary focus is on head motion because we are interested in the impact of head motion on the diffusion-weighted images used for tractography and the subsequent construction of structural connectivity. We provide additional analysis of the residual misalignment in Appendix G. This section explores how head motion estimates are obtained and its impact on brain structural connectomes.}

\textcolor{black}{We primarily rely on the FSL eddy tool to estimate head motion. The eddy software corrects head motion in diffusion MRI data through a multi-step process that involves modeling and correcting for both eddy currents and subject movement. It outputs a file named ``my\_eddy\_output.eddy\_movement\_rms,'' which provides summaries of the total movement in each volume. This is calculated by determining the displacement of each voxel, averaging the squares of those displacements across all intracerebral voxels, and finally taking the square root of that average. The file contains two columns: the first column lists the RMS (root mean square) movement relative to the first volume, and the second column lists the RMS movement relative to the previous volume. We use the second measure as our head motion metric in this paper (as it is the only head motion measure accessible in the ABCD data).}

\textcolor{black}{Figure \ref{fig:motion} (\textbf{A}) shows distributions of head motion for the ABCD and HCP studies. We notice that the motion distribution of ABCD subjects is more right-skewed with more subjects experiencing bigger motion during image acquisition.} This disparity in motion measure distributions between the ABCD and HCP studies could be attributed to the demographic differences in their respective cohorts. Our hypothesis suggests that the HCP study focuses on adults, who tend to follow instructions better and exhibit less movement during image acquisition compared to adolescents in the ABCD study.

To explore the impact of head motion on structural connectomes, we assign individuals to a large and a small motion group corresponding to each movement type. Figure \ref{fig:motion} shows the cutoff lines for both ABCD and HCP data. Specifically, in the ABCD data, individuals whose motion estimates are ranked among the top (bottom) 5 percent are assigned to the large (small) motion group. In the HCP data, we use the top and the bottom 10 percent as the threshold. We then quantify brain connection differences between groups. We subtract edge-specific means of the small motion group from those of the large motion group and show the results as adjacency matrices in Figure \ref{fig:motion} (\textbf{B}). In addition, we visualize between-group differences using a circular layout in Figure \ref{fig:motion} (\textbf{C}) to examine how connection patterns and strength differ across ROIs. Note that the dMRI data used here are preprocessed with the FSL eddy tool and its motion correction procedures. Figure \ref{fig:motion} (\textbf{B}, \textbf{C}) reveal that brain networks of large-motion subjects show systematic differences in fiber counts across multiple ROIs compared to those of small-motion subjects, implying that motion artifacts still affect structural connection reconstruction even after motion correction in the preprocessing stage \autocite{andersson2016a, andersson2016b}. Motivated by these findings, we propose the motion-invariant graph variational auto-encoder in Section \ref{sec:methods} to remove such systematic artifacts from the connectome data and downstream analyses.

\section{Method}\label{sec:methods}

The brain connectome for an individual is represented as an undirected graph $\mathcal{G} = (\mathcal{V}, \mathcal{E})$ with nodes $\mathcal{V}$ and edges $ \mathcal{E}$. 
The collection of nodes $\mathcal{V}$ refers to the partitioned brain regions, and the set of edges $\mathcal{E}$ represents connections between pairs of ROIs. The ROIs are pre-aligned so that each node in $\mathcal{V}$ corresponds to the same brain region across different subjects. We use the symmetric adjacency matrix $\boldsymbol{A}$ to represent $\mathcal{G}$ with $A_{uv}$ denoting the number of streamlines connecting ROIs $u$ and $v$; $A_{uu} = 0$ since we don't consider self-connections. For $N$ subjects, we denote the network data as $\boldsymbol{A}_i, i = 1, 2, \dots, N$. Also, denote the cognitive trait score for an individual as $y_i$, and the amount of motion as $\boldsymbol{c}_i$.  \textcolor{black}{Note that $\boldsymbol{c}_i$ can be a scalar quantifying overall head displacement or a vector ($\boldsymbol{c}_i \in \mathbb{R}^C$) with each dimension measuring different aspects of motion like the amplitudes of translation or rotation.} \textcolor{black}{Moreover, $\boldsymbol{c}_i$ can represent other undesirable artifacts, such as the residual misalignment, depending on the research goal.} 

We aim to infer the relationship between individuals' structural connectomes, recovered from neuroimaging data, and their measured cognitive traits, while adjusting for \textcolor{black}{motion-induced artifacts} during data acquisition. The key to achieving this goal is to learn a low-dimensional feature representation of \(\boldsymbol{A}_i\) as \(\boldsymbol{z}_i\) that is invariant to motion \(\boldsymbol{c}_i\), and then relate this feature \(\boldsymbol{z}_i\) to \(y_i\). Inspired by the variational auto-encoder (VAE) framework in \autocite{kingma2013}, we achieve our goal in three steps: 1) an encoder module (inference model) that learns the mapping from the observation \(\boldsymbol{A}_i\) to the motion-invariant latent variable \(\boldsymbol{z}_i\); 2) a decoder module (generative model) that specifies how to construct \(\boldsymbol{A}_i\) from \(\boldsymbol{z}_i\) and motion \(\boldsymbol{c}_i\); and 3) a regression module to link the motion-invariant \(\boldsymbol{z}_i\) to the cognitive trait \(y_i\); see Figure \ref{fig:model} for the model architecture. {\textcolor{black}{{Intuitively, in the training process, our approach seeks to maximize the joint likelihood of \((\boldsymbol{A}_i, y_i)\) conditional on $\boldsymbol{c}_i$  while minimizing the dependence between \(\boldsymbol{z}_i\) and \(\boldsymbol{c}_i\). This is achieved by characterizing the joint likelihood through the brain network generative model in Section \ref{sec:generative_model} and addressing the dependence through mutual information between the low-dimensional representations of brain networks and motion, as detailed in Section \ref{supervised_task}.}}

In the following sections, we start with introducing our graph generative model and then discuss the encoder and regression modules.

\subsection{Brain Network Generative Model}\label{sec:generative_model}
For each individual $i$, we assume the edges $A_{i[uv]}$ are conditionally independent given $\boldsymbol{z}_i$ and $\boldsymbol{c}_i$. The likelihood of the set of edges in $\boldsymbol{A}_i$ is  
\begin{equation}\label{eq:generative}
p_{\boldsymbol{\theta}}(\boldsymbol{A}_i \mid \boldsymbol{z}_i, \boldsymbol{c}_i) = \prod_{u=1}^{V} \prod_{v=1}^{V} p_{\boldsymbol{\theta}}(A_{i[uv]} \mid \boldsymbol{z}_i, \boldsymbol{c}_i),
\end{equation}
 where $p_{\boldsymbol{\theta}}(\boldsymbol{A}_i \mid \boldsymbol{z}_i, \boldsymbol{c}_i)$ is the generative model for $\boldsymbol{A}_i$ given $\boldsymbol{z}_i$ and $\boldsymbol{c}_i$, parameterized by  $\boldsymbol{\theta}$ learned via neural networks. We assume  $\boldsymbol{A}_i$ to be generated from the following process:
\begin{align}
\boldsymbol{z}_i &\sim \mathcal{N}(\boldsymbol{0},{\boldsymbol{I}}_K), \\ 
A_{i[uv]} \mid (\boldsymbol{z}_i, \boldsymbol{c}_i) &\sim   \; \textrm{Pois}(\lambda_{i[uv]}(\boldsymbol{z}_i, \boldsymbol{c}_i)), \label{eq:poisson}
\end{align}
where $K$ is the dimension of the latent feature space. The fiber count between ROIs $u$ and $v$ is modeled using a Poisson distribution with a parameter $\lambda_{i[uv]}(\boldsymbol{z}_i, \boldsymbol{c}_i)$, which relates to both the encoded feature $\boldsymbol{z}_i$ and motion $\boldsymbol{c}_i$. 

To capture the varying effect of motion across individuals and brain connections, we further model  $\lambda_{i[uv]}(\boldsymbol{z}_i, \boldsymbol{c}_i)$ as 
\begin{align}
\tilde{\boldsymbol{z}}_i &:= \boldsymbol{z}_i \oplus \boldsymbol{c}_i, \\
\lambda_{i[uv]}(\boldsymbol{z}_i, \boldsymbol{c}_i) &= \exp(\xi_{uv} + \psi_{uv}(\tilde{\boldsymbol{z}}_i)),
\label{eq:gamma} 
\end{align}
where $\oplus$ denotes the concatenation operator, and $\tilde{\boldsymbol{z}}_i \in \mathbb{R}^{K+C}$ is a concatenated vector representing motion-affected latent features. $\xi_{uv}$ is a baseline parameter controlling the population connection strength between regions $u$ and $v$, representing shared structural similarity across individuals, and $\psi_{uv}(\tilde{\boldsymbol{z}}_i)$ characterizes individual connection strength after accounting for motion. 

We model $\psi_{uv}(\tilde{\boldsymbol{z}}_i)$ based on the latent space model in \autocite{hoff2002}. 
The key idea is that the connection strength (fiber count) between regions $u$ and $v$ depends on their distance in some network latent space in $\mathbb{R}^R$. In this network latent space, each brain region is represented by a vector in the $\mathbb{R}^R$ space, with the distance between two points reflecting the degree of association between the corresponding brain regions. The closer two points are in this space, the stronger their connection is expected to be. The network latent space is different from the latent space of $\tilde{\boldsymbol{z}}_i$, which represents a low-dimensional space where the original high-dimensional brain structural connectomes are encoded into a lower-dimensional representation.

We construct a mapping $\bf{X}$ from $\mathbb{R}^{K+C}$ to $\mathbb{R}^{V\times R}$ that maps the motion-affected feature $\tilde{\boldsymbol{z}}_i$ to the network latent space that captures the latent positions of all brain regions, i.e., 
${\bf X}(\tilde{\boldsymbol{z}}_i)^\top = \left( {\bf X}_1(\tilde{\boldsymbol{z}}_i), \dots, {\bf X}_V(\tilde{\boldsymbol{z}}_i)\right)$ 
with $\boldsymbol{X}_u(\tilde{\boldsymbol{z}}_i) = ({X}_{u1}(\tilde{\boldsymbol{z}}_i),...,{X}_{uR}(\tilde{\boldsymbol{z}}_i))^\top\in \mathbb{R}^R$ for brain region $u$. 
Specifically, we assume each node $u\in \mathcal{V}$ for individual $i$ lies in a latent space  $\mathbb{R}^R$ and  represent $\psi_{uv}(\tilde{\boldsymbol{z}}_i)$ as 
\begin{align}
\psi_{uv}(\tilde{\boldsymbol{z}}_i) = \sum_{r=1}^R \alpha_r {X}_{ur}(\tilde{\boldsymbol{z}}_i){X}_{vr}(\tilde{\boldsymbol{z}}_i), \label{eq:svd1}
\end{align}
where $\alpha_r > 0$ is a weight controlling the importance of the $r$-th dimension. A large positive inner product between $\boldsymbol{X}_u(\tilde{\boldsymbol{z}}_i)$ and $\boldsymbol{X}_v(\tilde{\boldsymbol{z}}_i)$ implies a large fiber count between this ROI pair. Therefore, $\boldsymbol{X}_u(\tilde{\boldsymbol{z}}_i)$ incorporates the geometric collaborations among nodes, and broadcasts the impact of motion to all brain connections. \autocite{liu2019} proposed a graph convolutional network (GCN) to learn a nonlinear mapping from the encoded feature $\boldsymbol{z}_i$ to $(\boldsymbol{X}_1(\boldsymbol{z}_i),\dots, \boldsymbol{X}_V(\boldsymbol{z}_i)),$ leveraging on unique geometric features of the brain networks. The same procedure is adopted in this paper with the additional adjustment for motion $\boldsymbol{c}_i$, and we defer the detailed GCN framework to Appendix A.

\subsection{Motion Invariant Brain Network Encoding and Traits Prediction}\label{supervised_task}
We are interested in studying the relationship between brain structural networks and human traits,
correcting for \textcolor{black}{motion-related artifacts}. The key is to find an optimal encoder that produces $\boldsymbol{z}_i$ from $\boldsymbol{A}_i$ invariant to $\boldsymbol{c}_i$, and then formulate a regression of the trait $y_i$ with respect to the latent feature vector $\boldsymbol{z}_i$. We represent the inference model (encoder) as $q_{\boldsymbol{\phi}}(\boldsymbol{z}_i \mid \boldsymbol{A}_i)$, the generative model (decoder) as $p_{\boldsymbol{\theta}}(\boldsymbol{A}_i \mid \boldsymbol{z}_i, \boldsymbol{c}_i)$, and regression as $p_{\boldsymbol{\theta}}(y_i \mid \boldsymbol{z}_i)$. 
Our invariant encoding and trait prediction task is to find $q_{\boldsymbol{\phi}}(\boldsymbol{z}_i \mid \boldsymbol{A}_i)$ and $p_{\boldsymbol{\theta}}(\boldsymbol{A}_i \mid \boldsymbol{z}_i, \boldsymbol{c}_i)$ that maximize $\mathbb{E}_{\boldsymbol{A}_i, y_i, \boldsymbol{c}_i}[\log p_{\boldsymbol{\theta}}(\boldsymbol{A}_i, y_i \mid \boldsymbol{c}_i)]$, subject to $\boldsymbol{z}_i$ being independent of $\boldsymbol{c}_i$. Here $p_{\boldsymbol{\theta}} (\boldsymbol{A}_i, y_i \mid \boldsymbol{c}_i)$ is the joint likelihood of $\boldsymbol{A}_i, y_i$ conditional on $\boldsymbol{c}_i$. 
Since correlation is a limited notion of statistical dependence, which does not capture non-linear dependencies, we instead 
 use mutual information between $\boldsymbol{z}_i$ and $\boldsymbol{c}_i$ to quantify their dependency similar to \autocite{moyer2018}, and formulate the objective function as follows:
\begin{align}
    \mathcal{L}(\boldsymbol{A}_i,& y_i, \boldsymbol{c}_i; \boldsymbol{\theta}, \boldsymbol{\phi}) 
    = \mathbb{E}_{\boldsymbol{A}_i, y_i, \boldsymbol{c}_i} [\log p_{\boldsymbol{\theta}}(\boldsymbol{A}_i, y_i \mid \boldsymbol{c}_i) ]
    - \lambda I(\boldsymbol{z}_i, \boldsymbol{c}_i),
    \label{eq: sup_elbo}
\end{align}
where $I(\boldsymbol{z}_i, \boldsymbol{c}_i)$ is the mutual information between $\boldsymbol{z}_i$ and $\boldsymbol{c}_i$,  and $\lambda$ is a trade-off parameter between the joint likelihood and the mutual information. $\boldsymbol{\phi}$ are parameters of the encoder that are learned via neural networks. 

To simplify our model, we assume 1) the human trait $y_i$ and the brain connectivity $\boldsymbol{A}_i$
are conditionally independent given the latent representation $\boldsymbol{z}_i$ for individual $i$, and 2) only $\boldsymbol{A}_i$ is affected by $\boldsymbol{c}_i$ so $y_i$ is not affected. Assumption 2) indicates that motion happens randomly given the population considered and does not relate to the human traits (e.g., cognitive ability) considered. In certain scenarios, this assumption might not hold. For example, in the older population, people with mild cognitive impairment (MCI) tend to move more than normal controls \autocite{haller2014}, and if $y_i$ is an indicator of MCI, Assumption 2) does not hold here. Assumption 1) indicates that the encoded feature $\boldsymbol{z}_i$ contains all the information from $\boldsymbol{A}_i$ needed to predict $y_i$, and the residuals are independent of each other. Intuitively, we assume the motion $\boldsymbol{c}_i$ explains part of the variability in brain networks $\boldsymbol{A}_i$, but this part of variability in $\boldsymbol{A}_i$ does not relate to $y_i$.  Of course, we can modify our model to add $\boldsymbol{c}_i$ into the prediction of $y_i$ in the case that $\boldsymbol{c}_i$ is not independent of $y_i$. But it will make the final results difficult to interpret since we are mostly concerned with the relationship between $\boldsymbol{A}_i$ and $y_i$. 

Under Assumptions 1) and 2), we can express the conditional likelihood of  $\boldsymbol{A}_i, y_i$ given $\boldsymbol{c}_i$ as 
\begin{align}
    \log p_{\boldsymbol{\theta}}&(\boldsymbol{A}_i, y_i \mid \boldsymbol{c}_i) \geq \mathbb{E}_{\boldsymbol{z}_i \sim q_{\boldsymbol{\phi}}}[\log p_{\boldsymbol{\theta}}(\boldsymbol{A}_i \mid \boldsymbol{z}_i, \boldsymbol{c}_i) + \log p_{\boldsymbol{\theta}}(y_i \mid \boldsymbol{z}_i)] - KL[q_{\boldsymbol{\phi}}(\boldsymbol{z}_i \mid \boldsymbol{A}_i)\ ||\ p(\boldsymbol{z}_i)]. \label{eq:cond_joint}
\end{align}
The derivation is based on Jensen's inequality; see Appendix B. Intuitively, the lower bound of $\log p_{\boldsymbol{\theta}}(\boldsymbol{A}_i, y_i \mid \boldsymbol{c}_i)$ in (\ref{eq:cond_joint}) consists of three parts. The first term, $\log p_{\boldsymbol{\theta}}(\boldsymbol{A}_i \mid \boldsymbol{z}_i, \boldsymbol{c}_i)$, is a reconstruction error that utilizes log-likelihood to measure how well the generative model in (\ref{eq:poisson}) reconstructs $\boldsymbol{A}_i$. The second term, $\log p_{\boldsymbol{\theta}}(y_i \mid \boldsymbol{z}_i)$, measures the trait prediction accuracy. The third KL divergence term is a regularizer that pushes $q_{\boldsymbol{\phi}}(\boldsymbol{z}_i 
\mid \boldsymbol{A}_i)$ to be close
to its prior $\mathcal{N}(\boldsymbol{0}, \boldsymbol{I}_K)$ so that we can sample $\boldsymbol{z}_i$ easily. 

For the second term in the objective function (\ref{eq: sup_elbo}), we can upper bound  $I(\boldsymbol{z}_i, \boldsymbol{c}_i)$ as  
\begin{align}
    I(\boldsymbol{z}_i, \boldsymbol{c}_i) 
     &\leq \mathbb{E}_{\boldsymbol{A}_i}[KL[q_{\boldsymbol{\phi}}(\boldsymbol{z}_i \mid \boldsymbol{A}_i)\ ||\ q_{\boldsymbol{\phi}}(\boldsymbol{z}_i)]] - \mathbb{E}_{\boldsymbol{A}_i,  \boldsymbol{c}_i, \boldsymbol{z}_i \sim q_{\boldsymbol{\phi}}}[\log (\boldsymbol{A}_i\mid \boldsymbol{z}_i, \boldsymbol{c}_i)], \label{eq: mutual_info}
\end{align}
which consists of two parts: the KL divergence between  $q_{\boldsymbol{\phi}}(\boldsymbol{z}_i \mid \boldsymbol{A}_i)$ and its marginal $q_{\boldsymbol{\phi}}(\boldsymbol{z}_i)$ to ensure less variability across the input data $\boldsymbol{A}_i$, and the reconstruction error. See Appendix C for detailed derivations of (\ref{eq: mutual_info}). 

Combining the log-likelihood and the mutual information terms, our training objective in (\ref{eq: sup_elbo}) can be expressed as 
\begin{align}
    \mathcal{L}&(\boldsymbol{A}_i, y_i, \boldsymbol{c}_i; \boldsymbol{\theta}, \boldsymbol{\phi}) =
    \mathbb{E}_{\boldsymbol{A}_i, y_i, \boldsymbol{c}_i}[\log p_{\boldsymbol{\theta}}(\boldsymbol{A}_i, y_i \mid \boldsymbol{c}_i)] - \lambda I(\boldsymbol{z}_i, \boldsymbol{c}_i) \nonumber 
    \\
    & \geq - \,\mathbb{E}_{\boldsymbol{A}_i, y_i, \boldsymbol{c}_i} \big[ KL\big[q_{\boldsymbol{\phi}}(\boldsymbol{z}_i \mid \boldsymbol{A}_i)\ ||\ p(\boldsymbol{z}_i)\big] + \lambda KL\big[q_{\boldsymbol{\phi}}(\boldsymbol{z}_i \mid \boldsymbol{A}_i)\ ||\ q_{\boldsymbol{\phi}}(\boldsymbol{z}_i)\big] \, \label{eq:inv_obj} \\
    & -(1+\lambda)\mathbb{E}_{\boldsymbol{z}_i\sim q_{\boldsymbol{\phi}}}[\log p_{\boldsymbol{\theta}}(\boldsymbol{A}_i \mid \boldsymbol{z}_i, \boldsymbol{c}_i)] 
    + \log p_{\boldsymbol{\theta}}(y_i \mid \boldsymbol{z}_i) \, \big]. \nonumber
\end{align}
Our goal is to maximize  $\mathcal{L}(\boldsymbol{A}_i, y_i, \boldsymbol{c}_i; \boldsymbol{\theta}, \boldsymbol{\phi})$ through its lower bound (\ref{eq:inv_obj}). In practice, the expectation in (\ref{eq:inv_obj}) is intractable and we employ Monte Carlo approximation for computation; see Appendix D. We define the approximated objective as $\tilde{\mathcal{L}}(\boldsymbol{A}_i, y_i, \boldsymbol{c}_i; \boldsymbol{\theta}, \boldsymbol{\phi})$, and implement a stochastic variational Bayesian
Algorithm \ref{algorithm} to update the parameters using mini-batch training. 
For the regression module, we consider $p_{\boldsymbol{\theta}}(y_i \mid \boldsymbol{z}_i)$ as a univariate Gaussian, i.e.,  $y_i \sim \mathcal{N}(\boldsymbol{z}_i^\top \boldsymbol{\beta} + b, \sigma^2)$, where $\boldsymbol{\beta}, b, \sigma^2 \in \boldsymbol{\theta}$ are parameters to be learned. We list the model parameters and architecture in the Appendix E. 

\begin{algorithm}
\textbf{Input:} $\{ \boldsymbol{A}_i
\}_{i=1}^N, \{ y_i\}_{i=1}^N, \{ \boldsymbol{c}_i\}_{i=1}^N$. 
\begin{algorithmic}
\State  Initialize $\boldsymbol{\theta}, \boldsymbol{\phi}$
\While {not converged} \State Sample a batch of $\{\boldsymbol{A}_i\}_{i=1}^m$, denoted as $\boldsymbol{\mathcal{A}}_m$.
\ForAll{$ \boldsymbol{A}_i \in \boldsymbol{\mathcal{A}}_m$}
\State Sample $\epsilon_i \in \mathcal{N}(\boldsymbol{0}, \boldsymbol{I}_K)$. 
\State Compute $\boldsymbol{z}_i = \boldsymbol{\mu}_{\boldsymbol{\phi}}(\boldsymbol{A}_i) + \boldsymbol{\epsilon}_i \odot \boldsymbol{\Sigma}_{\boldsymbol{\phi}}(\boldsymbol{A}_i)$
\State where $\odot$ denotes the inner product operator.
\State Compute the gradients $\nabla_{\boldsymbol{\theta}}\mathcal{\tilde{L}}(\boldsymbol{A}_i, y_i, \boldsymbol{c}_i; \boldsymbol{\theta}, \boldsymbol{\phi})$ 
\State and $\nabla_{\boldsymbol{\phi}}\mathcal{\tilde{L}}(\boldsymbol{A}_i, y_i, \boldsymbol{c}_i; \boldsymbol{\theta}, \boldsymbol{\phi})$ with $\boldsymbol{z}_i$.
\EndFor
\State Average the gradients across the batch.
\State Update $\boldsymbol{\theta}, \boldsymbol{\phi}$ using gradients of $\boldsymbol{\theta}, \boldsymbol{\phi}$.
\EndWhile
\State Return $\boldsymbol{\theta}, \boldsymbol{\phi}$.
\caption{Motion-Invariant Variational Auto-Encoders}\label{algorithm}
\end{algorithmic}
\end{algorithm}

\section{Simulation Study}\label{sec:simulation}

We first conduct a simulation study to verify the efficacy of our method in removing unwanted information from the data. We simulate random graphs with the community structure \autocite{nowicki2001} using the Python package \textit{NetworkX}. Random graphs of two communities are considered: Nodes in the same group are connected with a probability of 0.25, and nodes of different groups are connected with a probability of 0.01. 
We simulate 1,000 such networks $\{\boldsymbol{A}_i\}_{i=1}^{n}$, where $\boldsymbol{A}_i \in \mathbb{R}^{V\times V}$ with $V = 68$ nodes and $n = 1,000$. Figure \ref{fig:simulation} (\textbf{A}) displays the average network across 1,000 simulated networks. Next, we simulate a \textit{nuisance} variable, $\{s_i\}_{i=1}^n$, by sampling $20\%$ of its elements from $\mathcal{N}(0.6, 0.05)$ and the rest from $\mathcal{N}(1, 0.01)$. We generate the nuisance-affected networks $\tilde{\boldsymbol{A}_i}$ by propagating $s_i$ across all edges in $\boldsymbol{A}_i$ as follows: 
\begin{align}
    \tilde{\boldsymbol{A}_i} = (s_i \boldsymbol{A}_i)^\top (s_i \boldsymbol{A}_i). \label{eq:sim}
\end{align}
Intuitively, $s_i \sim \mathcal{N}(0.6, 0.05)$ introduces large artifacts, since its multiplication with $\boldsymbol{A}_i$ in (\ref{eq:sim}) results in a $\tilde{\boldsymbol{A}_i}$ with reduced edge values; $s_i \sim \mathcal{N}(1, 0.01)$ generates small artifacts as the multiplication operation in (\ref{eq:sim}) changes the edges of $\boldsymbol{A}_i$ to a lesser extent. The average network across 1,000 such nuisance-affected networks is displayed in Figure \ref{fig:simulation} (\textbf{A}). To visualize the impact of the simulated nuisance variable, the edge-specific differences of the large and small nuisance groups are computed for both simulated networks $\{\boldsymbol{A}_i\}_{i=1}^n$ and nuisance-affected networks $\{\tilde{\boldsymbol{A}_i}\}_{i=1}^n$ (Figure \ref{fig:simulation} (\textbf{B})). We observe apparent differences in edge values between the large and small nuisance groups of $\{\tilde{\boldsymbol{A}_i}\}_{i=1}^n$. We further apply principal component analysis (PCA) to both $\{\boldsymbol{A}_i\}_{i=1}^n$ and $\{\tilde{\boldsymbol{A}_i}\}_{i=1}^n$, and visualize their projections onto the first two principal components (PCs) in Figure \ref{fig:simulation} (\textbf{C}). The first two PCs of $\{\boldsymbol{A}_i\}_{i=1}^n$ are indistinguishable. Note that these simulated networks are nuisance-free, even though we color their projections with the simulated nuisance for visualization purposes. On the contrary, we observe an apparent separation between the large and small nuisance groups of  $\{\tilde{\boldsymbol{A}_i}\}_{i=1}^n$ after introducing unwanted associations between the nuisance variable and the simulated networks. 

\begin{figure}[t]
\centering
\includegraphics[width=.9\linewidth]{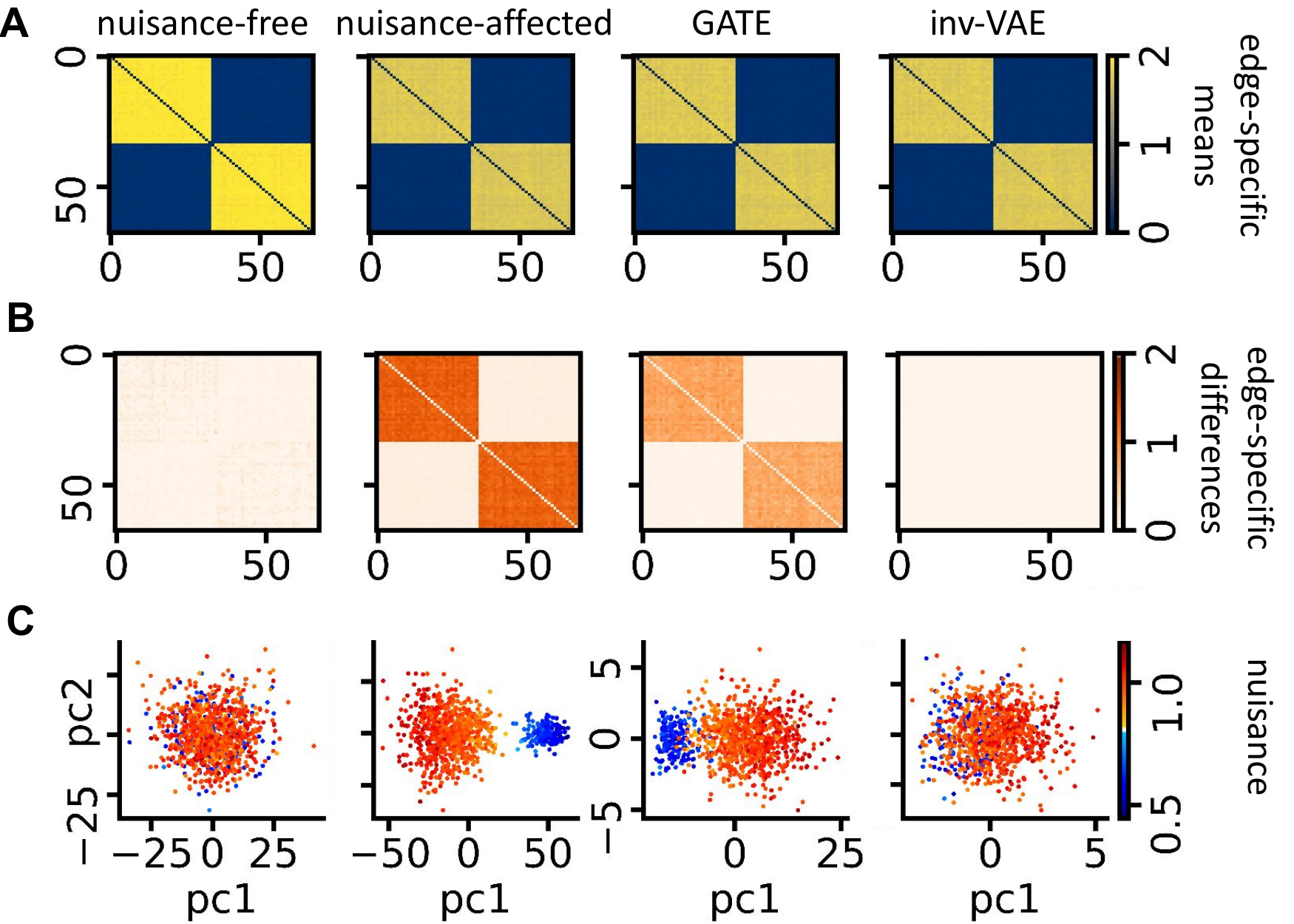}
\caption{(\textbf{A}) Edge-specific means of simulated networks, nuisance-affected networks, reconstructed networks by GATE and nuisance-corrected networks by inv-VAE. (\textbf{B}) Edge-specific differences between networks affected by large and small nuisance artifacts. The order of plots follows that in (\textbf{A}). (\textbf{C}) PCA projections of simulated networks, nuisance-affected networks, latent embeddings learned by GATE, and invariant embeddings from inv-VAE colored by the amount of nuisance artifacts.} \label{fig:simulation}
\end{figure}

To demonstrate the effectiveness of our method in removing artifacts introduced by the nuisance variable, we further compare inv-VAE with the graph auto-encoder (GATE) developed in \autocite{liu2019}. GATE has the same generative model as inv-VAE (see Section \ref{sec:methods})  except without adjusting for ${\bf c}_i$. Specifically, GATE assumes the following generative process for the brain network $\boldsymbol{A}_i$:
\begin{align}
    A_{i[uv]} \mid \boldsymbol{z}_i &\sim   \; \textrm{Poisson}\big(\lambda_{i[uv]}(\boldsymbol{z}_i)\big), \quad \boldsymbol{z}_i \sim \mathcal{N}(\boldsymbol{0},{\boldsymbol{I}}_K), \nonumber\\
    & \lambda_{i[uv]}(\boldsymbol{z}_i) = \exp(\xi_{uv} + \psi_{uv}(\boldsymbol{z}_i)), \nonumber\\ 
    & \psi_{uv}(\boldsymbol{z}_i) = \sum_{r=1}^R \alpha_r {X}_{ur}(\boldsymbol{z}_i){X}_{vr}(\boldsymbol{z}_i). \label{eq:gate_generative}
\end{align}Moreover, GATE learns the latent vector $\boldsymbol{z}_i$ using the following objective
    \begin{align}
    &\mathcal{L}(\boldsymbol{A}_i, y_i; \boldsymbol{\theta}, \boldsymbol{\phi}) = \mathbb{E}_{\boldsymbol{z}_i\sim q_{\boldsymbol{\phi}}}[\log p_{\boldsymbol{\theta}}(y_i \mid \boldsymbol{z}_i)] +  \mathbb{E}_{\boldsymbol{z}_i\sim q_{\boldsymbol{\phi}}}[\log p_{\boldsymbol{\theta}}(\boldsymbol{A}_i \mid \boldsymbol{z}_i)] 
    - KL[q_{\boldsymbol{\phi}}(\boldsymbol{z}_i \mid \boldsymbol{A}_i) \ ||\ p_{\boldsymbol{\theta}}(\boldsymbol{z}_i)]\label{eq:gate_objective}
    \end{align}
    instead of the motion-invariant objective in (\ref{eq:inv_obj}). We train GATE using $\{\tilde{\boldsymbol{A}_i}\}_{i=1}^n$ to learn latent representations of nuisance-affected networks. The learned latents are used by the generative model in (\ref{eq:gate_generative}) to reconstruct networks in Figure \ref{fig:simulation} (\textbf{A}). Figure \ref{fig:simulation} (\textbf{B}) and (\textbf{C}) show the reconstructed edge-specific differences between the large and small nuisance group, and the first two PCs of the learned latent variables, respectively.

Next, we train inv-VAE with $\{\tilde{\boldsymbol{A}_i}\}_{i=1}^n$ to learn parameters $\boldsymbol{\phi}$ and $\boldsymbol{\theta}$ for the inference and generative model. The inference model then learns motion-invariant latent representations, from which the generative model produces nuisance-corrected networks (Figure \ref{fig:simulation} (\textbf{A})) by setting $s_i = 1$. As previously defined, $s_i \approx 1$ corresponds to small nuisance artifacts, whereas a small $s_i $ represents  large nuisance artifacts. This is because multiplication of $\boldsymbol{A}_i$ with a small $s_i$ in (\ref{eq:sim}) shrinks its edge values to a large degree, whereas multiplication with $s_i \approx 1$ keeps the edge values roughly unchanged. 
Between-group edge-specific differences of the nuisance-corrected networks, and the first two PCs of invariant latents are shown in Figure \ref{fig:simulation} (\textbf{B}, \textbf{C}). Figure \ref{fig:simulation} (\textbf{B}, \textbf{C}) together suggest that GATE reconstructs the artifacts introduced by the nuisance variable, and its learned latent embeddings are affected by nuisance artifacts. On the contrary, such edge-specific differences between the large and small nuisance groups do not exist in the nuisance-corrected networks from inv-VAE, and projections of invariant embeddings of the large and small nuisance groups are indistinguishable. This observation implies that inv-VAE has the ability to remove unwanted information from the data.

\begin{figure}
\centering
\includegraphics[width=0.65\linewidth]{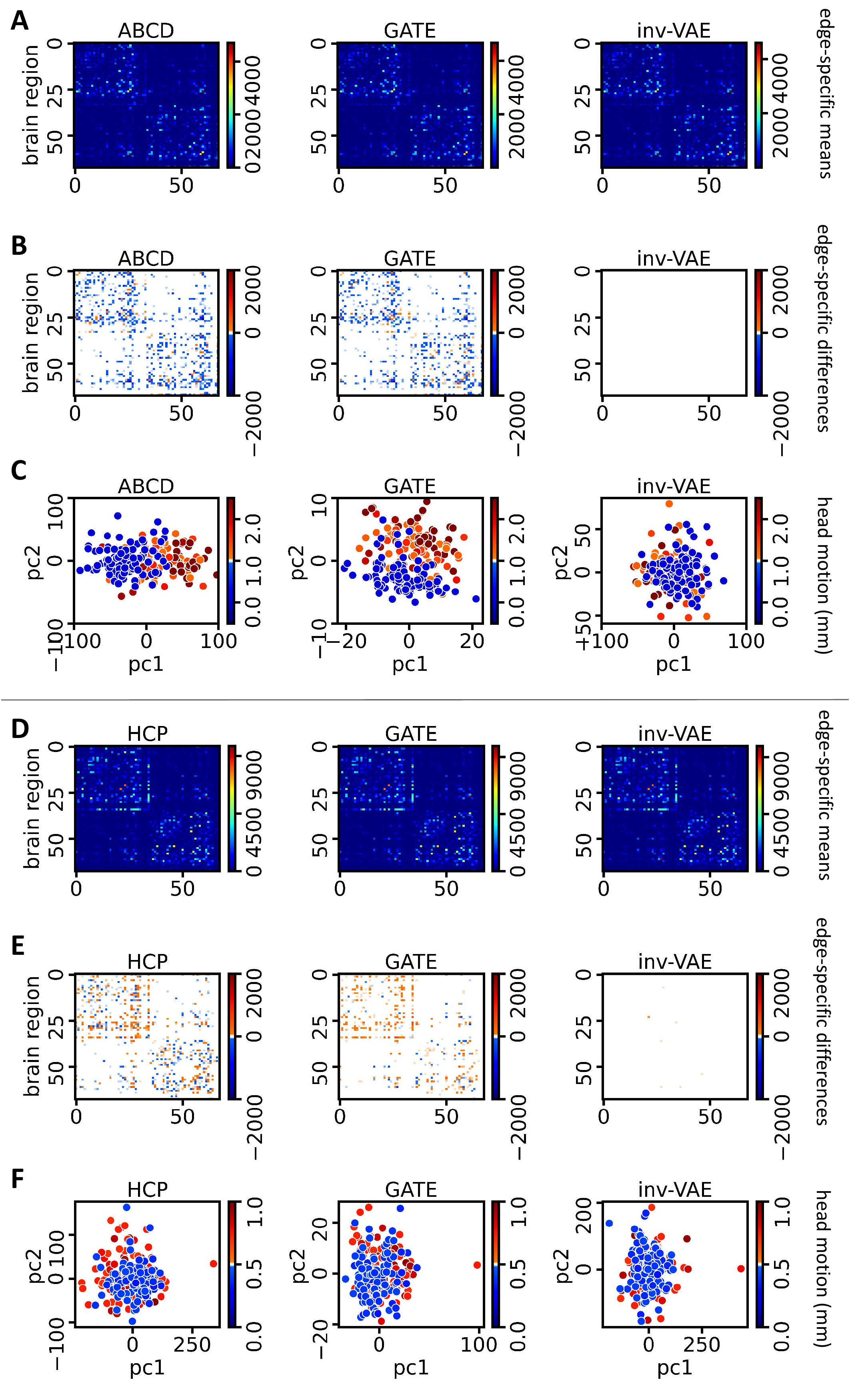}
\caption{(\textbf{A}) From left to right:  edge-specific means of ABCD brain networks, reconstructed brain networks by GATE, and motion-adjusted brain networks by inv-VAE, color-coded according to fiber count. (\textbf{B}) From left to right: edge-specific differences (large group subtracts small group) for raw network data, GATE reconstructed network data, and inv-VAE reconstructed data, respectively, with color indicating differences in fiber count. (\textbf{C}) From left to right: the first two principal component scores of ABCD brain networks, latent embeddings learned by GATE, and invariant embeddings from inv-VAE for observations in the large and small-motion groups, colored by the amount of head motion. (\textbf{D})-(\textbf{F}) replicate the same content as (\textbf{A})-(\textbf{C}) but emphasize head motion in the HCP data. \textcolor{black}{Head motion is quantified in millimeters (mm).}} \label{fig:  motion_adjust}
\end{figure}

\section{Applications to the ABCD and HCP Data}\label{sec:applications}

\subsection{Understanding and Removing Motion Artifacts}\label{sec:experiment_motion}
A major motivation of our work is to understand how motion affects learning a low-dimensional representation of the structural connectome and to remove \textcolor{black}{motion-induced} artifacts from the connectome data. 
For this purpose, we apply our inv-VAE to the ABCD (8,646 subjects) and HCP data (1065 subjects), and compare inv-VAE to its competitors on two tasks: (1) learning a latent connectome representation that is uninformative of motion; (2) generating a motion-adjusted connectome. \textcolor{black}{This section focuses on the FSL eddy head motion. We provide an additional analysis in Appendix G on addressing residual misalignment after eddy motion correction.}

We first average the \textit{motion-affected brain network} $\boldsymbol{A}_i$ over all individuals in the study to obtain the \textit{motion-affected edge-specific means} (see Section \ref{sec:motion_artifacts}) of the ABCD and HCP datasets, respectively. We call the resulting matrices as ``motion-affected'', meaning no motion adjustment has been done. The first columns of Figure \ref{fig: motion_adjust} (\textbf{A}, \textbf{D}) show the motion-affected edge-specific means in both datasets. We assign subjects into a small and a large motion group for each type of movement; See Section \ref{sec:motion_artifacts} for the cut-off thresholds for the large and small motion groups in both datasets.  
The first columns of Figure \ref{fig: motion_adjust} (\textbf{B}, \textbf{E}) show the \textit{motion-affected edge-specific differences}, obtained by subtracting the motion-affected edge-specific means of the small motion group from those of the large motion group. For both datasets, we notice apparent fiber count differences between the two motion groups in almost all pairs of brain regions. Such brain connection differences due to motion artifacts are more prominent when we visualize the first two PCs of brain networks between the two groups. In the first columns of Figure \ref{fig: motion_adjust} (\textbf{C}, \textbf{F}), each point represents an individual colored with the corresponding amount of motion. 
In the first column of Figure \ref{fig: motion_adjust} (\textbf{C}), the PCA projections of the large and small motion subjects are closer compared to those in the first column of Figure \ref{fig: motion_adjust} (\textbf{F}). This may be explained by the amount of \textcolor{black}{motion-induced} artifacts in the data: participants of the HCP study are young adults who may move less during data acquisition relative to the young adolescent ABCD participants. 

Next, we apply GATE \autocite{liu2019} to the ABCD and HCP data. For each dataset, GATE is trained using all brain networks $\boldsymbol{A}_i$'s to learn the model parameters $\boldsymbol{\theta}$ and $\boldsymbol{\phi}$. Equipped with well-trained estimators, GATE obtains the learned latent variables from the inference model and reconstructs brain networks from the generative model in (\ref{eq:gate_generative}). We denote the \textit{reconstructed brain network} as $\hat{\boldsymbol{A}_i}$. The second columns of Figure \ref{fig: motion_adjust} (\textbf{A}, \textbf{D}) show the \textit{reconstructed edge-specific means} averaged across all $\hat{\boldsymbol{A}_i}$'s in both datasets. In the second columns of Figure \ref{fig: motion_adjust} (\textbf{B}, \textbf{E}), \textit{reconstructed edge-specific differences} between the large and small motion group are shown. Although GATE can generate brain networks that resemble the observed data, motion artifacts are also undesirably reconstructed, as shown in the second columns of Figure \ref{fig: motion_adjust} (\textbf{B}, \textbf{E}). The second columns of Figure \ref{fig: motion_adjust} (\textbf{C}, \textbf{F}) show the first two PCs of the learned latents by GATE from the large and small motion group. The gap between the two groups of latents suggests that representations learned by GATE are also corrupted by motion artifacts, which will further affect prediction and inferences. 

\textcolor{black}{For our inv-VAE, we use $\boldsymbol{c}_i \in \mathbb{R}$ to represent the $i$-th subject's head motion, with 0 representing the smallest amount of motion.} To adjust for motion artifacts, we first train the inv-VAE model using the ABCD and HCP data separately. The inference model of inv-VAE learns a set of motion-invariant representations $\boldsymbol{z}_i$. Following the generative model in Section \ref{sec:generative_model}, 
$\boldsymbol{z}_i$ are subsequently used to generate \textit{motion-adjusted brain networks}, denoted as $\boldsymbol{A}^*_i$, by setting $\boldsymbol{c}_i$ to 0 to remove motion artifacts. For each dataset, we average across all $\boldsymbol{A}^*_i$ to obtain the \textit{motion-adjusted edge-specific means} shown in the third columns of Figure \ref{fig: motion_adjust} (\textbf{A}, \textbf{D}), which resemble the observed edge-specific means. 
The third columns of Figure \ref{fig: motion_adjust} (\textbf{B}, \textbf{E}) show the \textit{motion-adjusted edge-specific differences} by taking the difference between the motion-adjusted edge-specific means of the large and small motion group. Compared to the color scale of the observed between-group edge-specific differences, the colors of the motion-adjusted between-group edge-specific differences are noticeably lighter, suggesting that brain connection differences between the large and small motion groups become smaller after motion adjustment. The third columns of Figure \ref{fig: motion_adjust} (\textbf{C}, \textbf{F}) show the first two PCs of invariant representations $\boldsymbol{z}_i$ for the ABCD and HCP data. The gap between large and small motion subjects is smaller after motion-adjustment compared to the projections of the observed brain networks. Although the HCP subjects have smaller head movement and their connectomes are less affected by \textcolor{black}{motion-induced} artifacts, we still observe that motion-adjustment minimizes the reconstructed structural connection difference between the large and small motion group in the third column of Figure \ref{fig: motion_adjust} (\textbf{F}).  

\begin{figure}[t]
\centering
\includegraphics[width=\linewidth]{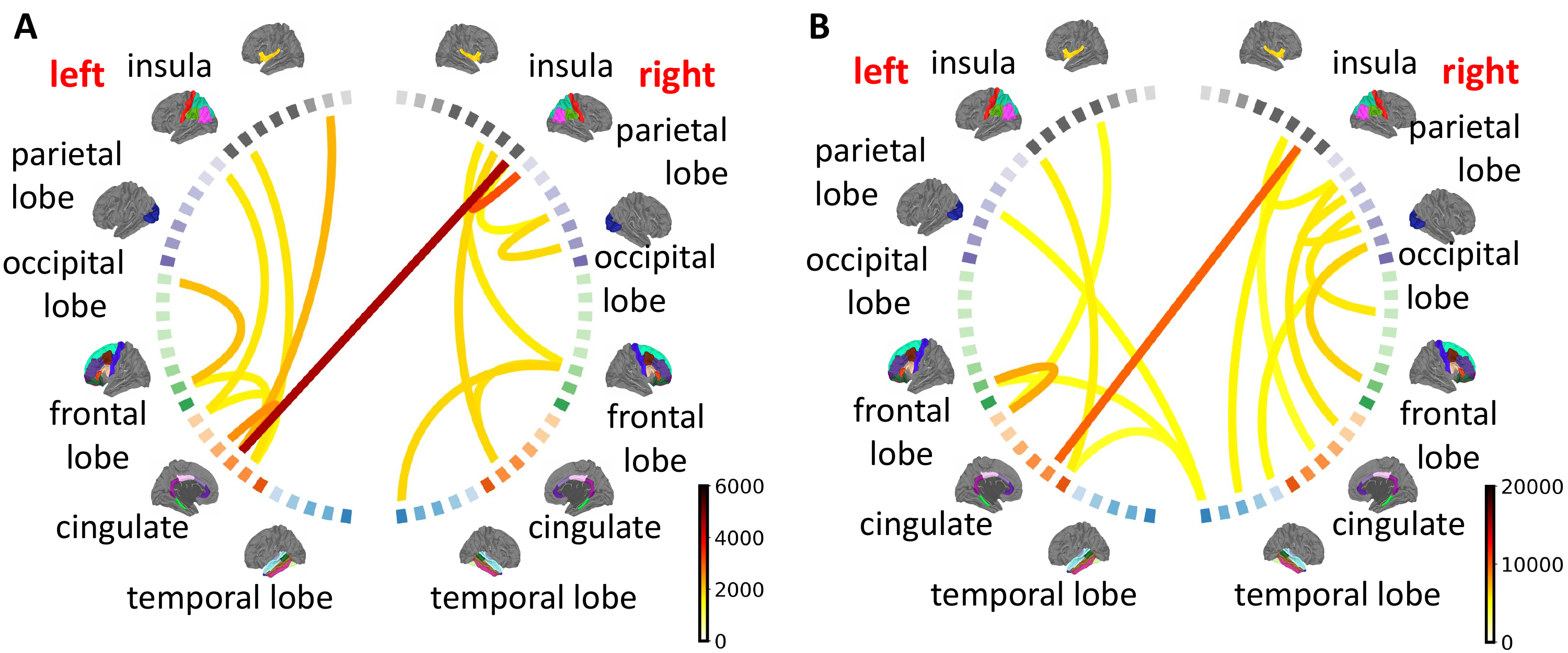}
\caption{Structural connectivity differences between motion-adjusted and motion-affected ABCD (\textbf{A}) and HCP (\textbf{B}) brain networks; see Section \ref{sec:experiment_motion} for definition. Each line connecting a pair of brain regions is colored by the corresponding fiber count difference after motion adjustment. We only show the 15 mostly changed brain connections. }\label{fig:circ_plots}
\end{figure}

\textcolor{black}{Motion can significantly impact the acquired diffusion data, introducing signal loss, geometric distortion, misalignment volumes, and compromised diffusion measures. There have been many studies to see how motion impacts the diffusion MRI signal and constructed structural connectomes. For example, \autocite{andersson2016b} highlight the susceptibility of diffusion MRI to artifacts caused by subject movements, resulting in both signal loss and geometric distortion. Additionally, \autocite{baum2018impact} demonstrate that motion has a noticeable impact on a considerable percentage of edges in structural brain networks. This effect is not limited to network edges but extends to significantly affected node strength and total network strength.} In this work, we aim to understand how motion impacts our understanding of structural brain connectivity.
For both ABCD and HCP studies, we first obtain the motion-affected edge-specific means via averaging $\boldsymbol{A}_i$'s across all individuals in the dataset. After training inv-VAE using all $\boldsymbol{A}_i$'s, we generate motion-adjusted networks $\boldsymbol{A}^*_i$'s from the  invariant embeddings. We then average across all $\boldsymbol{A}^*_i$'s to construct the motion-adjusted edge-specific means. The differences between the motion-adjusted and motion-affected edge-specific means are shown in Figure \ref{fig:circ_plots} (\textbf{A}, \textbf{B}). In both datasets, we observe more connections between the cingulate and the other brain regions in both hemispheres after motion adjustment, implying that brain connections related to the cingulate are more severely impacted by motion artifacts. There are also notable impacts in other regions, including the frontal, temporal, occipital, and parietal lobes, where we observe significant changes in connection strength after motion adjustment. 

\subsection{Relating Motion-Adjusted Connectomes to Cognition-Related Traits}\label{sec:experiment_trait}

Our method can effectively adjust for \textcolor{black}{motion-induced} artifacts in the connectome data. A natural question is whether motion-invariant representations further our understanding of the relationship between brain connectomes and cognition-related traits. From the ABCD dataset, we extract the following cognitive traits as $y_i$: (1) picture vocabulary test score for language comprehension; (2) oral reading recognition test score for language decoding; (3) fluid composite score for new learning; (4) crystallized composite score for past learning; (5) cognition total composite score for overall cognition capacity. From the HCP dataset, we consider similar traits:
(1) picture vocabulary test score; (2) pattern completion test score for processing speed; (3) picture sequence test score for episodic memory; (4) list sorting test score for working memory; (5) fluid intelligence score; see \autocite{gershon2013} for details about these traits. All traits are age-corrected.

According to Section \ref{supervised_task}, we assume that motion $c_i$ is independent from the trait of interest $y_i$. However, there might be significant correlations between clinically and scientifically interesting traits and subject motion. To mitigate the concern of erroneously attributing part of the connectome differences to motion rather than underlying traits, we pre-process the trait $y_i$ by regressing out $c_i$. For our analysis, we predict the residuals of $y_i$ after regressing out $c_i$ and examine the relationship between motion-adjusted connectomes and cognition-related traits.}

\begin{figure}[t]
\centering
\includegraphics[width=.9\linewidth]{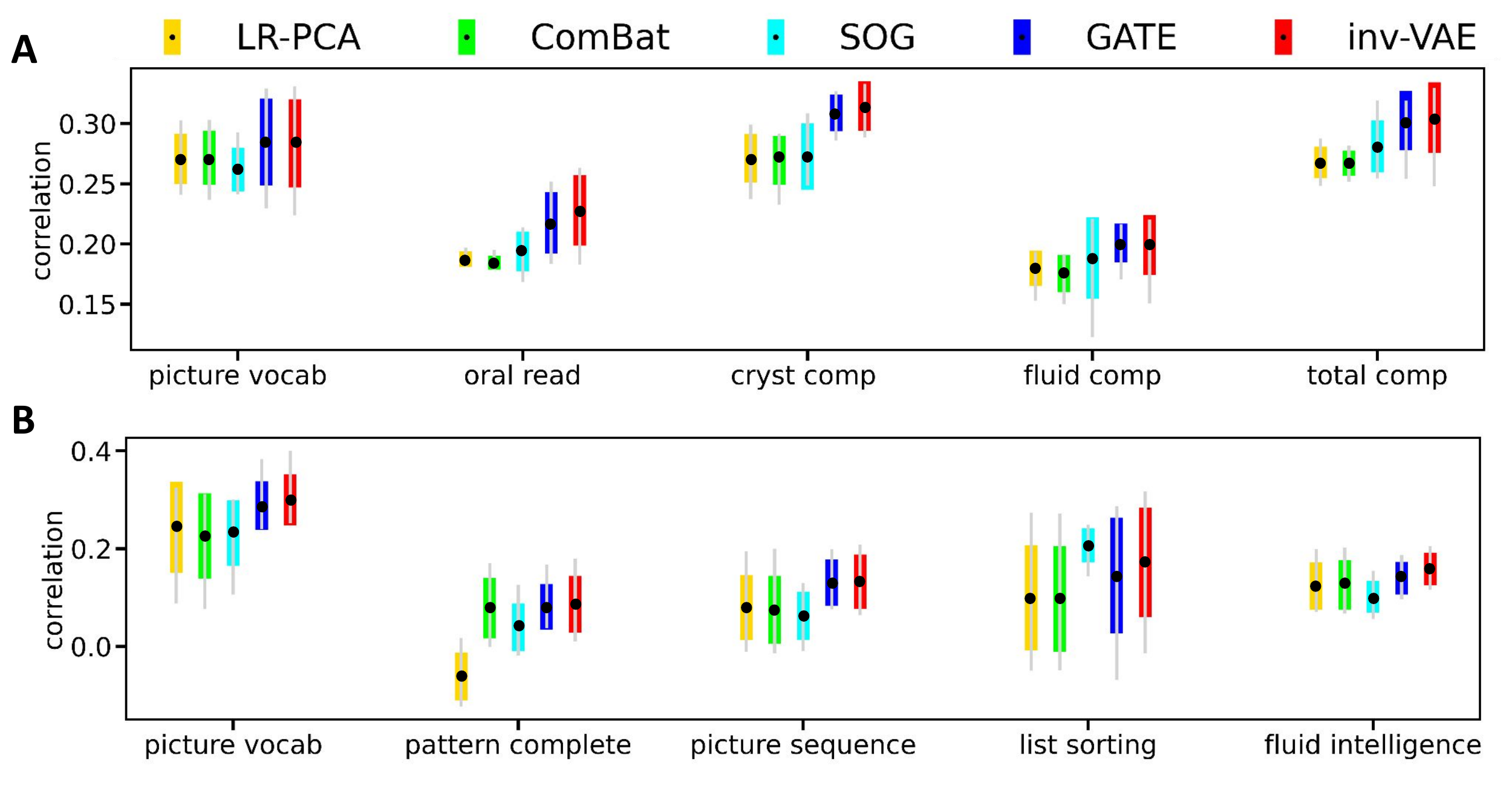}
\caption{A comparison of all methods on trait prediction in the ABCD (\textbf{A}) and HCP (\textbf{B}) data. The error bars show the min, max, mean, and standard deviation of correlation coefficients from 5-fold CV. We pre-process the target trait $y_i$ by regressing out head motion. For each fold, we train inv-VAE on the training data. For the test data, we obtain trait predictions after setting setting $c_i$ to 0 to remove head motion.} \label{fig:pred}
\end{figure}

To investigate whether our method outperforms other approaches in relating structural connectomes to traits, we compare inv-VAE to the following competitors: (1) \textit{LR-PCA} (motion-adjusted) that uses PCA to obtain a lower-dimensional projection from each individual's brain matrix, and then links $y_i$ to the projection via a linear regression (LR). Translation and rotation are covariates that LR adjusts for; (2) \textit{Sparse Orthogonal to Group (SOG)} \autocite{aliverti2021} that utilizes matrix decomposition to produce an adjusted dataset that is statistically independent of the motion variable. We then regress $y_i$ onto the obtained low-dimensional factorizations via LR; (3) \textit{ComBat} \autocite{johnson2007} that uses empirical Bayes for batch-effect removal, and returns a motion-adjusted dataset. We then apply PCA to the adjusted data to obtain low-dimensional projections as input to LR; (4) \textit{GATE} \autocite{liu2019} that has the same architecture as inv-VAE, but does not adjust for motion; see Section \ref{sec:simulation} for details about the model specification.

We use the above competing methods to 
produce 68-dimensional connectome representations for each individual, consistent with the recommended latent dimension $K$ in inv-VAE; see Section \ref{sec:methods}. To assess trait prediction quality, we use the Pearson correlation coefficient to measure the correlation between the observed and predicted traits, and performe 5-fold cross-validation (CV). A large correlation coefficient means the predicted trait scores more closely match the observed ones, indicating a stronger association between the low-dimensional representations of structural connectomes and the cognition-related traits. 

\begin{figure}[t]
\centering
\includegraphics[width=\linewidth]{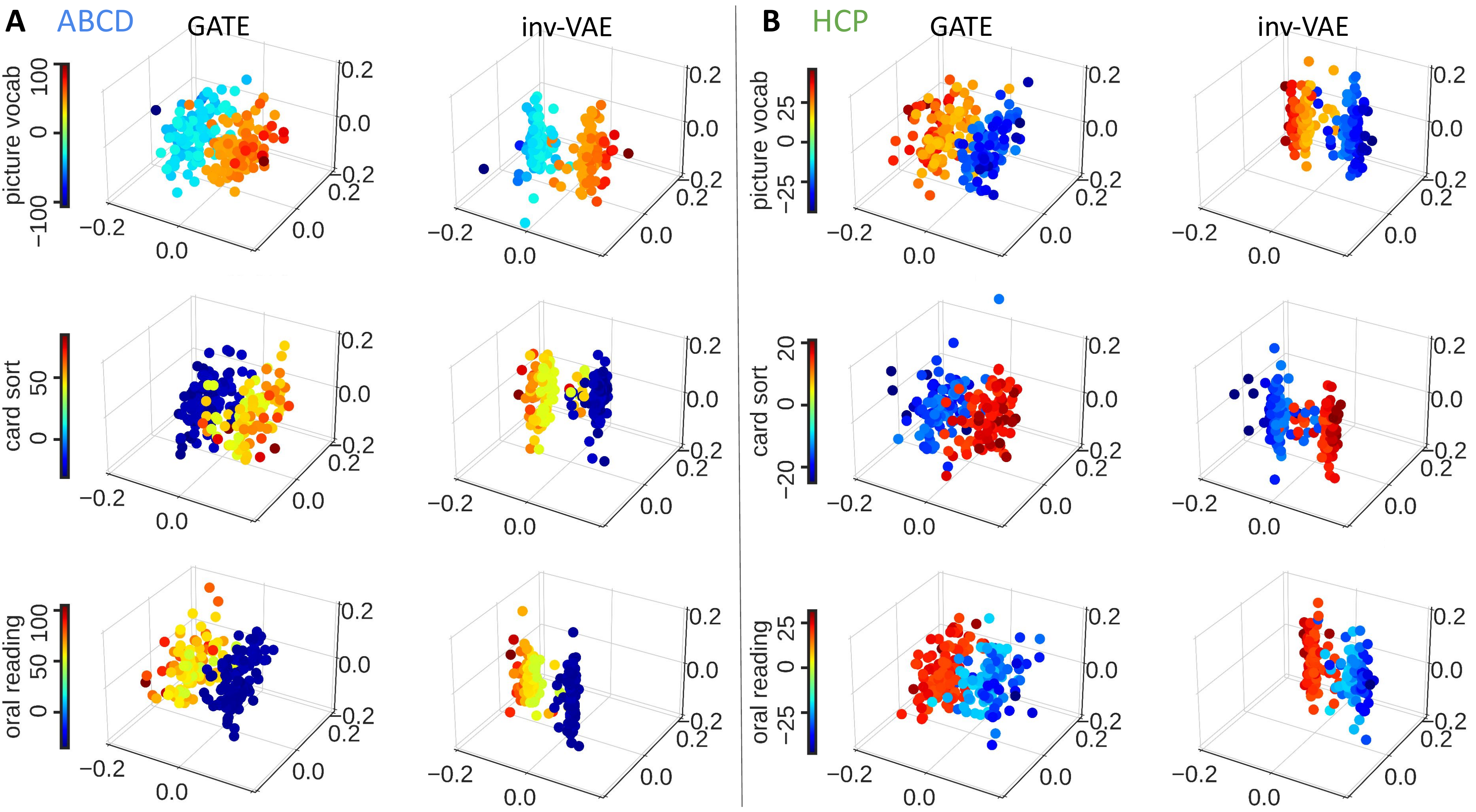}
\caption{Relationships between learned embeddings and cognition-related traits in ABCD (\textbf{A}) and HCP (\textbf{B}) datasets. {We pre-process the target trait $y_i$ by regressing out head motion. Both inv-VAE and GATE are trained with brain networks of all individuals in each dataset to obtain the latent $\boldsymbol{z}_i$'s for the $y_i$'s. For each trait, latent features from 200 subjects are selected, with the first group of 100 having the lowest trait scores and the second group of 100 having the highest scores. The first three PCs of the posterior means of $\boldsymbol{z}_i$ colored with their corresponding trait scores are displayed for each trait $y_i$.} Latent embeddings produced by GATE and motion-invariant embeddings from inv-VAE are compared.}  \label{fig:pca}
\end{figure}

A comparison of inv-VAE to the other approaches w.r.t. the trait-prediction quality is displayed in Figure \ref{fig:pred}, which shows that inv-VAE outperforms  LR-PCA, ComBat and SOG over all studied traits in both datasets, and either outperforms or is comparable to GATE over most selected traits. The prediction quality of SOG is comparable with GATE and inv-VAE for some cognitive traits: oral reading recognition and fluid intelligence score.  

Another way to assess the predictive performance is to visualize the associations between low-dimensional representations of structural connectomes and cognition-related traits. Both inv-VAE and GATE produce latent features $\boldsymbol{z}_i$ for each individual, which can be used for visualizations. Particularly, for the ABCD study, we consider picture vocabulary test,  oral reading recognition test and crystalized composite score; for the HCP study, we consider picture vocabulary test,  oral reading recognition test and dimensional change card sort test. Both inv-VAE and GATE are trained with brain networks of all individuals in each dataset to obtain the $\boldsymbol{z}_i$'s for the $y_i$'s. For each trait, latent features from 200 subjects are selected, with the first group of 100 having the lowest trait scores and the second group of 100 having the highest scores. Next, we visualize the first three PCs of the posterior means of $\boldsymbol{z}_i$ colored with their corresponding trait scores in Figure \ref{fig:pca}. For both inv-VAE and GATE, Figure \ref{fig:pca} shows that representations from the two trait groups are separated, suggesting that structural connectomes are different among individuals with different cognitive abilities. We highlight that motion-invariant $\boldsymbol{z}_i$'s of the two groups are more separated than motion-affected $\boldsymbol{z}_i$'s from GATE, particularly for the  oral reading recognition test. It implies that motion-invariant structural connectomes are more strongly associated with cognition-related traits than other approaches without motion adjustment.

\section{Discussion}\label{sec:discussion}
We develop a motion invariant variational auto-encoder (inv-VAE) for learning low-dimensional, motion-adjusted representations of structural connectomes. We apply inv-VAE to the Adolescent Brain Cognitive Development (ABCD) and Human Connectome Project (HCP) data and discover noticeable motion artifacts in both, despite the incorporation of motion correction procedures in preprocessing. This observation reinforces the need for effective motion mitigation strategies in connectome analysis. Being invariant to motion artifacts during the connectome modeling phase, inv-VAE shows improved performance in understanding the correlation between structural connectomes and cognition-related traits. While our inv-VAE is motivated to handle the motion confounder, it can be easily extended to handle other confounders, such as site batch effect, variation due to machine settings, etc. Compared with popular batch effect removing models such as ComBat, the advantages of our inv-VAE include 1) it can handle non-linear objects (brain network data), and 2) it can deal with non-linear batch effects.  Therefore, we believe that our inv-VAE framework can be extended to broader brain imaging analysis tasks, highlighting the importance of confounder or batch effect adjustment methodologies.

Note that inv-VAE is developed under a generative model framework, which allows us to simulate brain networks under various conditions. The simulation of brain networks is crucial since brain network data are generally not widely available, and the extraction of brain networks from raw imaging data is far from straightforward. Our inv-VAE enables us to simulate $(A_i, c_i, y_i)$ — that is, the brain network, \textcolor{black}{undesirable} artifacts, and cognition ability measures — simultaneously. Such simulated data can be freely shared with biostatisticians and data scientists who wish to develop statistical models for brain networks, but prefer not to get involved in the intricate process of brain imaging preprocessing. 

Our method can model the spatially heterogeneous impact of motion: edges that share the same node can have correlations under our model since they are functions of both $z_i$ and $c_i$. In this work, we simplify $c_i$ to encode only a summary of subject-level motion artifacts. One possible future direction is to explore the representations of motion that can improve our generative model.

Our proposed inv-VAE currently models the motion-affected structural connectome using a Poisson generative model in conjunction with a GCN. Future work could incorporate pre-existing neuroscience knowledge regarding the effects of \textcolor{black}{motion-induced} artifacts on specific brain regions or connections to refine our model and improve the reconstruction of motion-adjusted structural connectomes. This could be achieved by leveraging empirical evidence, neuroscientific findings, or theoretical frameworks that shed light on the disparate impacts of motion on individual brain regions or connections. Through this integration of knowledge, we aim to boost the precision and accuracy of our motion artifact mitigation process, enabling us better to capture the subtle influences of motion on brain connectivity.

Another future direction is to leverage recent developments of generative modeling frameworks,  such as variational diffusion models \autocite{kingma2021variational}, score-based generative models \autocite{song2020improved} and Hierarchical Variational Autoencoders (HVAE) \autocite{sonderby2016ladder}. These models can be understood under the general Evidence Lower Bound (ELBO) method used here in this paper but with different constraints in the latent space \autocite{luo2022understanding}.  Moreover, we use \textcolor{black}{the mean motion estimate across frames, as computed by the FSL eddy tool,} to represent motion during diffusion MRI acquisition and ignore the longitudinal or higher-order information in the motion. This simplification may not fully  represent the impacts of motion on diffusion MRI data, and thus the motion-adjusted connectomes from our inv-VAE may not be completely invariant to motion. Future developments may incorporate more complex summaries of dynamic motion.

\vspace{+.8cm}
\noindent \textbf{ACKNOWLEDGMENTS}\\
\noindent
This work is supported by grants from the National Science Foundation (NSF-DMS-2124535).

\noindent \textbf{ETHICS STATEMENT}\\
\noindent The datasets were anonymized and not collected by the investigators; in which case the work is classified as nonhuman research.

\noindent \textbf{DATA AND CODE AVAILABILITY}\\
\noindent The data used in this study contains  publicly available datasets from the Human Connectome Project\footnote{\href{https://www.humanconnectome.org/}{https://www.humanconnectome.org/}} and the Adolescent Brain Cognitive Development Study\footnote{\href{https://nida.nih.gov/research-topics/adolescent-brain/longitudinal-study-adolescent-brain-cognitive-development-abcd-study}{https://nida.nih.gov/research-topics/adolescent-brain/longitudinal-study-adolescent-brain-cognitive-development-abcd-study}}. The model training and testing code is available at \href{https://github.com/yzhang511/inv-vae}{https://github.com/yzhang511/inv-vae}.

\noindent \textbf{AUTHOR CONTRIBUTIONS}\\
\noindent Conceptualization and Study Design: Yizi Zhang, Meimei Liu, Zhengwu Zhang. Data curation: Zhengwu Zhang. Methodology, Formal Analysis, Visualizations: Yizi Zhang, Meimei Liu. Manuscript
Writing (original draft): Yizi Zhang. Manuscript Writing (review \& editing): Yizi Zhang, Meimei Liu, Zhengwu Zhang, David Dunson. Supervision: David Dunson. All authors have substantively revised the
work, reviewed the manuscript, approved the submitted version, and agreed to be personally accountable for their contributions.

\noindent \textbf{DECLARATION OF COMPETING INTEREST}\\
\noindent The authors declare no competing interests.

\printbibliography{}

@article{glasser2013,
  title={The human connectome project's neuroimaging approach},
  author={Glasser, Matthew F and Smith, Stephen M and Marcus, Daniel S and Andersson, Jesper LR and Auerbach, Edward J and Behrens, Timothy EJ and Coalson, Timothy S and Harms, Michael P and Jenkinson, Mark and Moeller, Steen and others},
  journal={Nature neuroscience},
  volume={19},
  number={9},
  pages={1175--1187},
  year={2016},
  publisher={Nature Publishing Group US New York},
}

@article{bjork2017,
  title={The ABCD study of neurodevelopment: Identifying neurocircuit targets for prevention and treatment of adolescent substance abuse},
  author={Bjork, James M and Straub, Lisa K and Provost, Rosellen G and Neale, Michael C},
  journal={Current treatment options in psychiatry},
  volume={4},
  pages={196--209},
  year={2017},
  publisher={Springer}
}

@article{zhang2018a,
  title={Mapping population-based structural connectomes},
  author={Zhang, Zhengwu and Descoteaux, Maxime and Zhang, Jingwen and Girard, Gabriel and Chamberland, Maxime and Dunson, David and Srivastava, Anuj and Zhu, Hongtu},
  journal={NeuroImage},
  volume={172},
  pages={130--145},
  year={2018},
  publisher={Elsevier}
}

@article{zhang2019,
  title={Tensor network factorizations: Relationships between brain structural connectomes and traits},
  author={Zhang, Zhengwu and Allen, Genevera I and Zhu, Hongtu and Dunson, David},
  journal={Neuroimage},
  volume={197},
  pages={330--343},
  year={2019},
  publisher={Elsevier}
}

@article{andersson2016a,
  title={Incorporating outlier detection and replacement into a non-parametric framework for movement and distortion correction of diffusion MR images},
  author={Andersson, Jesper LR and Graham, Mark S and Zsoldos, Enik{\H{o}} and Sotiropoulos, Stamatios N},
  journal={Neuroimage},
  volume={141},
  pages={556--572},
  year={2016},
  publisher={Elsevier}
}

@inproceedings{zemel2013,
  title={Learning fair representations},
  author={Zemel, Rich and Wu, Yu and Swersky, Kevin and Pitassi, Toni and Dwork, Cynthia},
  booktitle={International conference on machine learning},
  pages={325--333},
  year={2013},
  organization={PMLR}
}

@article{achille2018,
  title={Emergence of invariance and disentanglement in deep representations},
  author={Achille, Alessandro and Soatto, Stefano},
  journal={Journal of Machine Learning Research},
  volume={19},
  number={50},
  pages={1--34},
  year={2018}
}

@article{kochunov2006,
  title={Retrospective motion correction protocol for high-resolution anatomical MRI},
  author={Kochunov, Peter and Lancaster, Jack L and Glahn, David C and Purdy, David and Laird, Angela R and Gao, Feng and Fox, Peter},
  journal={Human brain mapping},
  volume={27},
  number={12},
  pages={957--962},
  year={2006},
  publisher={Wiley Online Library}
}

@article{bhushan2015,
  title={Co-registration and distortion correction of diffusion and anatomical images based on inverse contrast normalization},
  author={Bhushan, Chitresh and Haldar, Justin P and Choi, Soyoung and Joshi, Anand A and Shattuck, David W and Leahy, Richard M},
  journal={Neuroimage},
  volume={115},
  pages={269--280},
  year={2015},
  publisher={Elsevier}
}

@article{ben-amitay2012,
  title={Motion correction and registration of high b-value diffusion weighted images},
  author={Ben-Amitay, Shani and Jones, Derek K and Assaf, Yaniv},
  journal={Magnetic resonance in medicine},
  volume={67},
  number={6},
  pages={1694--1702},
  year={2012},
  publisher={Wiley Online Library}
}

@article{alter2000,
  title={Singular value decomposition for genome-wide expression data processing and modeling},
  author={Alter, Orly and Brown, Patrick O and Botstein, David},
  journal={Proceedings of the National Academy of Sciences},
  volume={97},
  number={18},
  pages={10101--10106},
  year={2000},
  publisher={National Acad Sciences}
}

@article{aliverti2021,
  title={Removing the influence of group variables in high-dimensional predictive modelling},
  author={Aliverti, Emanuele and Lum, Kristian and Johndrow, James E and Dunson, David B},
  journal={Journal of the Royal Statistical Society Series A: Statistics in Society},
  volume={184},
  number={3},
  pages={791--811},
  year={2021},
  publisher={Oxford University Press}
}

@article{benito2004,
  title={Adjustment of systematic microarray data biases},
  author={Benito, Monica and Parker, Joel and Du, Quan and Wu, Junyuan and Xiang, Dong and Perou, Charles M and Marron, James Stephen},
  journal={Bioinformatics},
  volume={20},
  number={1},
  pages={105--114},
  year={2004},
  publisher={Oxford University Press}
}

@article{johnson2007,
  title={Adjusting batch effects in microarray expression data using empirical Bayes methods},
  author={Johnson, W Evan and Li, Cheng and Rabinovic, Ariel},
  journal={Biostatistics},
  volume={8},
  number={1},
  pages={118--127},
  year={2007},
  publisher={Oxford University Press}
}

@article{moyer2018,
  title={Invariant representations without adversarial training},
  author={Moyer, Daniel and Gao, Shuyang and Brekelmans, Rob and Galstyan, Aram and Ver Steeg, Greg},
  journal={Advances in neural information processing systems},
  volume={31},
  year={2018}
}

@article{andersson2016b,
  title={An integrated approach to correction for off-resonance effects and subject movement in diffusion MR imaging},
  author={Andersson, Jesper LR and Sotiropoulos, Stamatios N},
  journal={Neuroimage},
  volume={125},
  pages={1063--1078},
  year={2016},
  publisher={Elsevier}
}

@article{casey2018,
  title={The adolescent brain cognitive development (ABCD) study: imaging acquisition across 21 sites},
  author={Casey, Betty Jo and Cannonier, Tariq and Conley, May I and Cohen, Alexandra O and Barch, Deanna M and Heitzeg, Mary M and Soules, Mary E and Teslovich, Theresa and Dellarco, Danielle V and Garavan, Hugh and others},
  journal={Developmental cognitive neuroscience},
  volume={32},
  pages={43--54},
  year={2018},
  publisher={Elsevier}
}

@article{essen2012,
  title={The Human Connectome Project: a data acquisition perspective},
  author={Van Essen, David C and Ugurbil, Kamil and Auerbach, Edward and Barch, Deanna and Behrens, Timothy EJ and Bucholz, Richard and Chang, Acer and Chen, Liyong and Corbetta, Maurizio and Curtiss, Sandra W and others},
  journal={Neuroimage},
  volume={62},
  number={4},
  pages={2222--2231},
  year={2012},
  publisher={Elsevier}
}

@article{maier2017,
  title={The challenge of mapping the human connectome based on diffusion tractography},
  author={Maier-Hein, Klaus H and Neher, Peter F and Houde, Jean-Christophe and C{\^o}t{\'e}, Marc-Alexandre and Garyfallidis, Eleftherios and Zhong, Jidan and Chamberland, Maxime and Yeh, Fang-Cheng and Lin, Ying-Chia and Ji, Qing and others},
  journal={Nature communications},
  volume={8},
  number={1},
  pages={1349},
  year={2017},
  publisher={Nature Publishing Group UK London}
}

@article{desikan2006,
  title={An automated labeling system for subdividing the human cerebral cortex on MRI scans into gyral based regions of interest},
  author={Desikan, Rahul S and S{\'e}gonne, Florent and Fischl, Bruce and Quinn, Brian T and Dickerson, Bradford C and Blacker, Deborah and Buckner, Randy L and Dale, Anders M and Maguire, R Paul and Hyman, Bradley T and others},
  journal={Neuroimage},
  volume={31},
  number={3},
  pages={968--980},
  year={2006},
  publisher={Elsevier}
}

@article{iannopollo2019,
  title={Surface-based analysis of cortical thickness and volume loss in Alzheimer’s disease},
  author={Iannopollo, Emily and Plunkett, Ryan and Garcia, Kara},
  journal={Proceedings of IMPRS},
  volume={2},
  number={1},
  year={2019}
}

@article{fornito2013,
  title={Graph analysis of the human connectome: promise, progress, and pitfalls},
  author={Fornito, Alex and Zalesky, Andrew and Breakspear, Michael},
  journal={Neuroimage},
  volume={80},
  pages={426--444},
  year={2013},
  publisher={Elsevier}
}

@article{gershon2013,
  title={NIH toolbox for assessment of neurological and behavioral function},
  author={Gershon, Richard C and Wagster, Molly V and Hendrie, Hugh C and Fox, Nathan A and Cook, Karon F and Nowinski, Cindy J},
  journal={Neurology},
  volume={80},
  number={11\_supplement\_3},
  pages={S2--S6},
  year={2013},
  publisher={AAN Enterprises}
}

@article{kingma2013,
  title={Auto-encoding variational bayes},
  author={Kingma, Diederik P},
  journal={arXiv preprint arXiv:1312.6114},
  year={2013}
}

@article{hoff2002,
  title={Latent space approaches to social network analysis},
  author={Hoff, Peter D and Raftery, Adrian E and Handcock, Mark S},
  journal={Journal of the american Statistical association},
  volume={97},
  number={460},
  pages={1090--1098},
  year={2002},
  publisher={Taylor \& Francis}
}

@article{liu2019,
  title={Graph auto-encoding brain networks with applications to analyzing large-scale brain imaging datasets},
  author={Liu, Meimei and Zhang, Zhengwu and Dunson, David B},
  journal={Neuroimage},
  volume={245},
  pages={118750},
  year={2021},
  publisher={Elsevier}
}

@article{haller2014,
  title={Head motion parameters in fMRI differ between patients with mild cognitive impairment and Alzheimer disease versus elderly control subjects},
  author={Haller, Sven and Monsch, Andreas U and Richiardi, Jonas and Barkhof, Frederik and Kressig, Reto W and Radue, Ernst W},
  journal={Brain topography},
  volume={27},
  pages={801--807},
  year={2014},
  publisher={Springer}
}

@article{nowicki2001,
  title={Estimation and prediction for stochastic blockstructures},
  author={Nowicki, Krzysztof and Snijders, Tom A B},
  journal={Journal of the American statistical association},
  volume={96},
  number={455},
  pages={1077--1087},
  year={2001},
  publisher={Taylor \& Francis}
}

@article{louizos2015,
  title={The variational fair autoencoder},
  author={Louizos, Christos and Swersky, Kevin and Li, Yujia and Welling, Max and Zemel, Richard},
  journal={arXiv preprint arXiv:1511.00830},
  year={2015}
}

@article{kingma2014,
  title={Adam: A method for stochastic optimization},
  author={Kingma, Diederik P},
  journal={arXiv preprint arXiv:1412.6980},
  year={2014}
}

@article{sonderby2016ladder,
  title={Ladder variational autoencoders},
  author={S{\o}nderby, Casper Kaae and Raiko, Tapani and Maal{\o}e, Lars and S{\o}nderby, S{\o}ren Kaae and Winther, Ole},
  journal={Advances in neural information processing systems},
  volume={29},
  year={2016}
}

@article{kingma2021variational,
  title={Variational diffusion models},
  author={Kingma, Diederik and Salimans, Tim and Poole, Ben and Ho, Jonathan},
  journal={Advances in neural information processing systems},
  volume={34},
  pages={21696--21707},
  year={2021}
}

@article{song2020improved,
  title={Improved techniques for training score-based generative models},
  author={Song, Yang and Ermon, Stefano},
  journal={Advances in neural information processing systems},
  volume={33},
  pages={12438--12448},
  year={2020}
}

@article{luo2022understanding,
  title={Understanding diffusion models: A unified perspective},
  author={Luo, Calvin},
  journal={arXiv preprint arXiv:2208.11970},
  year={2022}
}

@article{goodfellow2020generative,
  title={Generative adversarial networks: An overview},
  author={Creswell, Antonia and White, Tom and Dumoulin, Vincent and Arulkumaran, Kai and Sengupta, Biswa and Bharath, Anil A},
  journal={IEEE signal processing magazine},
  volume={35},
  number={1},
  pages={53--65},
  year={2018},
  publisher={IEEE}
}

@inproceedings{papernot2016limitations,
  title={The limitations of deep learning in adversarial settings},
  author={Papernot, Nicolas and McDaniel, Patrick and Jha, Somesh and Fredrikson, Matt and Celik, Z Berkay and Swami, Ananthram},
  booktitle={2016 IEEE European symposium on security and privacy (EuroS\&P)},
  pages={372--387},
  year={2016},
  organization={IEEE}
}

@inproceedings{chen2020simple,
  title={A simple framework for contrastive learning of visual representations},
  author={Chen, Ting and Kornblith, Simon and Norouzi, Mohammad and Hinton, Geoffrey},
  booktitle={International conference on machine learning},
  pages={1597--1607},
  year={2020},
  organization={PMLR}
}

@inproceedings{he2020momentum,
  title={Momentum contrast for unsupervised visual representation learning},
  author={He, Kaiming and Fan, Haoqi and Wu, Yuxin and Xie, Saining and Girshick, Ross},
  booktitle={Proceedings of the IEEE/CVF conference on computer vision and pattern recognition},
  pages={9729--9738},
  year={2020}
}

@article{khosla2020supervised,
  title={Supervised contrastive learning},
  author={Khosla, Prannay and Teterwak, Piotr and Wang, Chen and Sarna, Aaron and Tian, Yonglong and Isola, Phillip and Maschinot, Aaron and Liu, Ce and Krishnan, Dilip},
  journal={Advances in neural information processing systems},
  volume={33},
  pages={18661--18673},
  year={2020}
}

@article{aliee2023conditionally,
  title={Conditionally Invariant Representation Learning for Disentangling Cellular Heterogeneity},
  author={Aliee, Hananeh and Kapl, Ferdinand and Hediyeh-Zadeh, Soroor and Theis, Fabian J},
  journal={arXiv preprint arXiv:2307.00558},
  year={2023}
}

@inproceedings{greenfeld2020robust,
  title={Robust learning with the hilbert-schmidt independence criterion},
  author={Greenfeld, Daniel and Shalit, Uri},
  booktitle={International Conference on Machine Learning},
  pages={3759--3768},
  year={2020},
  organization={PMLR}
}

@inproceedings{yu2021measuring,
  title={Measuring dependence with matrix-based entropy functional},
  author={Yu, Shujian and Alesiani, Francesco and Yu, Xi and Jenssen, Robert and Principe, Jose},
  booktitle={Proceedings of the AAAI Conference on Artificial Intelligence},
  volume={35},
  number={12},
  pages={10781--10789},
  year={2021}
}

@inproceedings{liu2021lsmi,
  title={Lsmi-sinkhorn: Semi-supervised mutual information estimation with optimal transport},
  author={Liu, Yanbin and Yamada, Makoto and Tsai, Yao-Hung Hubert and Le, Tam and Salakhutdinov, Ruslan and Yang, Yi},
  booktitle={Machine Learning and Knowledge Discovery in Databases. Research Track: European Conference, ECML PKDD 2021, Bilbao, Spain, September 13--17, 2021, Proceedings, Part I 21},
  pages={655--670},
  year={2021},
  organization={Springer}
}

@article{alemi2016deep,
  title={Deep variational information bottleneck},
  author={Alemi, Alexander A and Fischer, Ian and Dillon, Joshua V and Murphy, Kevin},
  journal={arXiv preprint arXiv:1612.00410},
  year={2016}
}

@article{cieslak2021qsiprep,
  title={QSIPrep: an integrative platform for preprocessing and reconstructing diffusion MRI data},
  author={Cieslak, Matthew and Cook, Philip A and He, Xiaosong and Yeh, Fang-Cheng and Dhollander, Thijs and Adebimpe, Azeez and Aguirre, Geoffrey K and Bassett, Danielle S and Betzel, Richard F and Bourque, Josiane and others},
  journal={Nature methods},
  volume={18},
  number={7},
  pages={775--778},
  year={2021},
  publisher={Nature Publishing Group US New York}
}

@article{christiaens2021scattered,
  title={Scattered slice SHARD reconstruction for motion correction in multi-shell diffusion MRI},
  author={Christiaens, Daan and Cordero-Grande, Lucilio and Pietsch, Maximilian and Hutter, Jana and Price, Anthony N and Hughes, Emer J and Vecchiato, Katy and Deprez, Maria and Edwards, A David and Hajnal, Joseph V and others},
  journal={Neuroimage},
  volume={225},
  pages={117437},
  year={2021},
  publisher={Elsevier}
}

@article{baum2018impact,
  title={The impact of in-scanner head motion on structural connectivity derived from diffusion MRI},
  author={Baum, Graham L and Roalf, David R and Cook, Philip A and Ciric, Rastko and Rosen, Adon FG and Xia, Cedric and Elliott, Mark A and Ruparel, Kosha and Verma, Ragini and Tun{\c{c}}, Birkan and others},
  journal={Neuroimage},
  volume={173},
  pages={275--286},
  year={2018},
  publisher={Elsevier}
}

@article{zhang2019nonparametric,
  title={Nonparametric bayes models of fiber curves connecting brain regions},
  author={Zhang, Zhengwu and Descoteaux, Maxime and Dunson, David B},
  journal={Journal of the American Statistical Association},
  year={2019},
  publisher={Taylor \& Francis}
}

@article{lebihan2006artifacts,
  title={Artifacts and pitfalls in diffusion MRI},
  author={Le Bihan, Denis and Poupon, Cyril and Amadon, Alexis and Lethimonnier, Franck},
  journal={Journal of Magnetic Resonance Imaging: An Official Journal of the International Society for Magnetic Resonance in Medicine},
  volume={24},
  number={3},
  pages={478--488},
  year={2006},
  publisher={Wiley Online Library}
}

@article{yendiki2014spurious,
  title={Spurious group differences due to head motion in a diffusion MRI study},
  author={Yendiki, Anastasia and Koldewyn, Kami and Kakunoori, Sita and Kanwisher, Nancy and Fischl, Bruce},
  journal={Neuroimage},
  volume={88},
  pages={79--90},
  year={2014},
  publisher={Elsevier}
}

\newpage
\appendix

\section*{Appendix}
\textcolor{black}{This section provides model details and additional numerical studies. The appendix is organized as follows:
\begin{itemize}
    \item Section A: Specification of GCN details used in both inv-VAE and GATE. 
    \item Section B: Derivation of the joint likelihood of brain networks $\textbf{A}_i$ and cognition traits $y_i$ conditional on the nuisance factor $\textbf{c}_i$ as defined in Equation (\ref{eq:cond_joint}). 
    \item Section C: Derivation of the mutual information between latent variable $\boldsymbol{z}_i$ and $\boldsymbol{c}_i$ as defined in Equation (\ref{eq: sup_elbo}).
    \item Section D and E: Implementation details, including Monte Carlo approximation of the training objective and model architecture details of our developed inv-VAE. 
    \item Section F: Investigation of correlation between motion score and outlier measure to assess the relationship between head motion, residual misalignment, and outlier data. 
    \item Section G: Analysis of the effect of residual misalignment adjustment on brain connectome reconstruction and trait prediction. 
    \item Section H: Expanded simulation study to evaluate our method's robustness in handling extreme motion-induced artifacts. 
    \item Section I: Methods for extracting residual misregistration representations.
\end{itemize}
}

\subsection*{\normalfont \textbf{A. GCN model specification}}
To exploit the local collaborative patterns among brain regions, \autocite{liu2019} propose to learn each region's representation by propagating node-specific $k$-nearest neighbor information. We use the relative distance between a pair of brain regions to represent their locality. The relative distance is measured through the length of the white matter fiber tract connecting the pair, and stored in a matrix $\boldsymbol{D} \in \mathbb{R}^{V \times V}$. $D_{uv}$ denotes the fiber length between regions $u$ and $v$. For $u$, its $k$-nearest neighbors (kNNs) can be defined according to $\boldsymbol{D}$. Graph convolution is defined to learn the $r$-th latent coordinate $\boldsymbol{X}_r(\boldsymbol{\tilde{z}}_i)$ for subject $i$. In particular, an $m$-layer GCN is  $\boldsymbol{X}^{(m)}_{r}(\boldsymbol{\tilde{z}}_i) = h_m(\boldsymbol{W}^{(m)}_r\boldsymbol{X}^{(m-1)}_{r}(\boldsymbol{\tilde{z}}_i) + b_m),$  for $2 \leq m \leq M$, with $\boldsymbol{X}^{(1)}_{r}(\boldsymbol{\tilde{z}}_i) = h_1(\boldsymbol{W}^{(1)}_r\boldsymbol{\tilde{z}}_i + b_1)$. 
$\boldsymbol{X}^{(m)}_{r}(\boldsymbol{\tilde{z}}_i)$ denotes the output of the $m$-th GCN layer, $h_m(\cdot)$ is an activation function of the $m$-th layer, and $\boldsymbol{W}^{(m)}_r$ is a weight matrix characterizing the convolution operation at the $m$-th layer. The embedding feature of each region at the $m$-th layer is determined by the weighted sum of itself and its nearest neighbor regions at the $(m - 1)$-th layer. When $m = 1$, $\boldsymbol{W}^{(1)}_r \in \mathbb{R}^{V \times (K+C)}$ maps $\boldsymbol{\tilde{z}}_i \in \mathbb{R}^{K+C}$ to the latent space $\boldsymbol{X}^{(1)}_{r}(\boldsymbol{\tilde{z}}_i) \in \mathbb{R}^V$. When $m \geq 2$, $\boldsymbol{W}^{(m)}_r$ is a $V \times V$ matrix with the $u$-th row $\boldsymbol{w}_{u\cdot}^{(r,m)}$ satisfying $w_{uv}^{(r,m)} > 0$ if $v = u$ or $v \in k_{r}\text{NN}(u)$, and $w_{uv}^{(r,m)} = 0$ otherwise. Here $k_{r}\text{NN}(u)$ denotes the $k$ nearest neighbors of region $u$ in the $r$-th dimension of the latent space.

\subsection*{\normalfont \textbf{B. Derivation of the joint likelihood of $\boldsymbol{A}_i, y_i$ conditional on $\boldsymbol{c}_i$ in  (\ref{eq:cond_joint})}}
The joint likelihood of $\boldsymbol{A}_i$ and $y_i$ given $\boldsymbol{c}_i$ is
\begin{align*}
    \log p_{\boldsymbol{\theta}}(\boldsymbol{A}_i, y_i \mid \boldsymbol{c}_i)
    &= \log \big(\int 
    \frac{p_{\boldsymbol{\theta}}(\boldsymbol{A}_i, y_i, \boldsymbol{z}_i \mid \boldsymbol{c}_i)}{q_{\boldsymbol{\phi}}(\boldsymbol{z}_i \mid \boldsymbol{A}_i)}q_{\boldsymbol{\phi}}(\boldsymbol{z}_i \mid \boldsymbol{A}_i)d\boldsymbol{z}_i \big) = \log \big( \mathbb{E}_{\boldsymbol{z}_i \sim q_{\boldsymbol{\phi}}}\big[\frac{p_{\boldsymbol{\theta}}(\boldsymbol{A}_i, y_i, \boldsymbol{z}_i \mid \boldsymbol{c}_i)}{q_{\boldsymbol{\phi}}(\boldsymbol{z}_i \mid \boldsymbol{A}_i)}\big]\big). 
\end{align*}
By Jensen's inequality, 
\begin{align*}
    \log &\big(E_{\boldsymbol{z}_i \sim q_{\boldsymbol{\phi}}}\big[\frac{p_{\boldsymbol{\theta}}(\boldsymbol{A}_i, y_i, \boldsymbol{z}_i \mid \boldsymbol{c}_i)}{q_{\boldsymbol{\phi}}(\boldsymbol{z}_i \mid \boldsymbol{A}_i)}\big]\big) 
    \geq \mathbb{E}_{\boldsymbol{z}_i \sim q_{\boldsymbol{\phi}}}[\log p_{\boldsymbol{\theta}}(\boldsymbol{A}_i, y_i, \boldsymbol{z}_i \mid \boldsymbol{c}_i) - \log q_{\boldsymbol{\phi}}(\boldsymbol{z}_i \mid \boldsymbol{A}_i)] \\
    &= \mathbb{E}_{\boldsymbol{z}_i \sim q_{\boldsymbol{\phi}}}[\log p_{\boldsymbol{\theta}}(\boldsymbol{A}_i, y_i \mid \boldsymbol{z}_i, \boldsymbol{c}_i) +\log p_{\boldsymbol{\theta}}(\boldsymbol{z}_i \mid \boldsymbol{c}_i) - \log q_{\boldsymbol{\phi}}(\boldsymbol{z}_i \mid \boldsymbol{A}_i)]. 
\end{align*}
Assuming that $\boldsymbol{z}_i$ is independent of $\boldsymbol{c}_i$, we have $p_{\boldsymbol{\theta}}(\boldsymbol{z}_i \mid \boldsymbol{c}_i) = p(\boldsymbol{z}_i)$. We then have
\begin{align*}
\mathbb{E}_{\boldsymbol{z}_i \sim q_{\boldsymbol{\phi}}}[\log p_{\boldsymbol{\theta}}(\boldsymbol{A}_i, y_i \mid \boldsymbol{z}_i, \boldsymbol{c}_i)] &+ \mathbb{E}_{\boldsymbol{z}_i \sim q_{\boldsymbol{\phi}}}[\log p_{\boldsymbol{\theta}}(\boldsymbol{z}_i) - \log q_{\boldsymbol{\phi}}(\boldsymbol{z}_i \mid \boldsymbol{A}_i)] \\
&= \mathbb{E}_{\boldsymbol{z}_i \sim q_{\boldsymbol{\phi}}}[\log p_{\boldsymbol{\theta}}(\boldsymbol{A}_i, y_i \mid \boldsymbol{z}_i, \boldsymbol{c}_i)] - KL[q_{\boldsymbol{\phi}}(\boldsymbol{z}_i \mid \boldsymbol{A}_i)\ ||\ p(\boldsymbol{z}_i)].
\end{align*}

\subsection*{\normalfont \textbf{C. Derivation of  mutual information $I(\boldsymbol{z}_i, \boldsymbol{c}_i)$ defined in (\ref{eq: sup_elbo})}}
From mutual information properties, we have that
\begin{align*}
    I(\boldsymbol{z}_i, \boldsymbol{c}_i) = I(\boldsymbol{z}_i, \boldsymbol{A}_i) - I(\boldsymbol{z}_i, \boldsymbol{A}_i\mid \boldsymbol{c}_i) + I(\boldsymbol{z}_i, \boldsymbol{c}_i \mid \boldsymbol{A}_i),
\end{align*}
where $
    I(\boldsymbol{z}_i, \boldsymbol{c}_i \mid \boldsymbol{A}_i) = H(\boldsymbol{z}_i \mid \boldsymbol{A}_i) - H(\boldsymbol{z}_i \mid \boldsymbol{A}_i, \boldsymbol{c}_i) = H(\boldsymbol{z}_i \mid \boldsymbol{A}_i) - H(\boldsymbol{z}_i \mid \boldsymbol{A}_i) = 0,
$ 
because the distribution of $\boldsymbol{z}_i$ solely
depends on $\boldsymbol{A}_i$ not on $\boldsymbol{c}_i$. Using mutual information properties, we write $I(\boldsymbol{z}_i, \boldsymbol{c}_i)$ as
\begin{align*}
    I(\boldsymbol{z}_i, \boldsymbol{c}_i) 
    & = I(\boldsymbol{z}_i, \boldsymbol{A}_i) - I(\boldsymbol{z}_i, \boldsymbol{A}_i \mid \boldsymbol{c}_i) = I(\boldsymbol{z}_i,  \boldsymbol{A}_i) - H(\boldsymbol{A}_i \mid \boldsymbol{c}_i) + H(\boldsymbol{A}_i\mid \boldsymbol{z}_i, \boldsymbol{c}_i).
\end{align*}
Using variational inequality, we have
\begin{align*}
    I(\boldsymbol{z}_i, \boldsymbol{c}_i)
    & \leq I(\boldsymbol{z}_i, \boldsymbol{A}_i) - H(\boldsymbol{A}_i\mid \boldsymbol{c}_i) - 
    \mathbb{E}_{\boldsymbol{A}_i, \boldsymbol{z}_i, \boldsymbol{c}_i \sim q_{\boldsymbol{\phi}}}[\log p_{\boldsymbol{\theta}}(\boldsymbol{A}_i \mid \boldsymbol{z}_i, \boldsymbol{c}_i)] \\
    & = \mathbb{E}_{\boldsymbol{z}_i, \boldsymbol{A}_i}[\log q_{\boldsymbol{\phi}}(\boldsymbol{z}_i \mid \boldsymbol{A}_i) - \log q_{\boldsymbol{\phi}}(\boldsymbol{z}_i)] - H(\boldsymbol{A}_i \mid \boldsymbol{c}_i) - \mathbb{E}_{\boldsymbol{A}_i, \boldsymbol{c}_i, \boldsymbol{z}_i \sim q_{\boldsymbol{\phi}}}[\log p_{\boldsymbol{\theta}}(\boldsymbol{A}_i \mid \boldsymbol{z}_i, \boldsymbol{c}_i)] \\
    & = \mathbb{E}_{\boldsymbol{A}_i}[KL[q_{\boldsymbol{\phi}}(\boldsymbol{z}_i \mid \boldsymbol{A}_i)\ ||\ q_{\boldsymbol{\phi}}(\boldsymbol{z}_i)]] - H(\boldsymbol{A}_i\mid \boldsymbol{c}_i) - \mathbb{E}_{\boldsymbol{A}_i, \boldsymbol{c}_i, \boldsymbol{z}_i \sim q_{\boldsymbol{\phi}}}[\log p_{\boldsymbol{\theta}}(\boldsymbol{A}_i \mid \boldsymbol{z}_i, \boldsymbol{c}_i)],  
\end{align*}
where $H(\boldsymbol{A}_i\mid \boldsymbol{c}_i)$ is a constant that can be ignored.

\subsection*{\normalfont \textbf{D. Monte Carlo approximation of inv-VAE objective}}
We use Monte Carlo variational inference \autocite{kingma2013} to approximate the intractable expectation in our invariant objective. Using reparametrization trick, for each $\ell = 1, \dots, L$, we sample 
    ${\boldsymbol{z}_i}^{\ell} \sim q_{\boldsymbol{\phi}}(\boldsymbol{z}_i \mid \boldsymbol{A}_i) = \mathcal{N}(\boldsymbol{\mu}_{\boldsymbol{\phi}}(\boldsymbol{A}_i), \boldsymbol{\Sigma}_{\boldsymbol{\phi}}(\boldsymbol{A}_i))$, where $\boldsymbol{\mu}_{\boldsymbol{\phi}}(\boldsymbol{A}_i) \in \mathbb{R}^K$ and $\boldsymbol{\Sigma}_{\boldsymbol{\phi}}(\boldsymbol{A}_i) \in \mathbb{R}^{K\times K}$. With ${\boldsymbol{z}_i}^{\ell}$, we decompose the  objective in (\ref{eq:inv_obj}) into three parts: 1) The reconstruction error
    \begin{align*}
    \mathcal{L}_{\text{recon}} = \mathbb{E}_{\boldsymbol{z}_i \sim q_{\boldsymbol{\phi}}}[\log p_{\boldsymbol{\theta}}(\boldsymbol{A}_i \mid \boldsymbol{z}_i, \boldsymbol{c}_i) + \log p_{\boldsymbol{\theta}}(y_i \mid \boldsymbol{z}_i)]
    \end{align*}
    can be approximated by 
$
    \tilde{\mathcal{L}}_{\text{recon}} = \frac{1}{L}\sum_{\ell=1}^{L}\left(\log p_{\boldsymbol{\theta}}(\boldsymbol{A}_i \mid \boldsymbol{z}_i^{\ell}, \boldsymbol{c}_i) + \log p_{\boldsymbol{\theta}}(y_i \mid \boldsymbol{z}_i^{\ell}) \right).$
    2) Because both $q_{\boldsymbol{\phi}}(\boldsymbol{z}_i \mid \boldsymbol{A}_i)$ and $p_{\boldsymbol{\theta}}(\boldsymbol{z}_i)$ are normally distributed, the KL divergence between them, 
    $$\mathcal{L}_{\text{prior}} = KL\big[q_{\boldsymbol{\phi}}(\boldsymbol{z}_i \mid \boldsymbol{A}_i)\ ||\ p(\boldsymbol{z}_i)\big],$$ can be approximated by
    $\tilde{\mathcal{L}}_{\text{prior}} = \frac{1}{2}\sum_{k=1}^{K}(\mu_k^2 + \sigma_k^2 -1 - \log(\sigma_k)^2),$ where $\mu_k, \sigma_k$ are the $k$-th element of $\boldsymbol{\mu}_{\boldsymbol{\phi}}(\boldsymbol{A}_i)$ and $\boldsymbol{\Sigma}_{\boldsymbol{\phi}}(\boldsymbol{A}_i)$. 
    3) $q_{\boldsymbol{\phi}}(\boldsymbol{z}_i)$ is the marginal distribution of $q_{\boldsymbol{\phi}}(\boldsymbol{z}_i \mid \boldsymbol{A}_i)$, and they are two Gaussians. We can approximate  
    $$\mathcal{L}_{\text{marg}} = KL[q_{\boldsymbol{\phi}}(\boldsymbol{z}_i \mid \boldsymbol{A}_i)\ ||\ q_{\boldsymbol{\phi}}(\boldsymbol{z}_i)]
    $$ with the following pairwise Gaussian KL divergence \autocite{louizos2015}:
    \begin{align*}
    &\tilde{\mathcal{L}}_{\text{marg}} 
    = KL[q_{\boldsymbol{\phi}}(\boldsymbol{z}_i \mid \boldsymbol{A}_i)\ ||\ q_{\boldsymbol{\phi}}(\boldsymbol{z}_i)] 
    = \frac{1}{2} \big[\log \frac{\det(\boldsymbol{\Sigma}_0)}{\det(\boldsymbol{\Sigma}_1)} + \text{tr}(\boldsymbol{\Sigma}_0^{-1}\boldsymbol{\Sigma}_1) 
    - K + (\boldsymbol{\mu}_0 - \boldsymbol{\mu}_1)^\top \boldsymbol{\Sigma}_1^{-1} (\boldsymbol{\mu}_0 - \boldsymbol{\mu}_1) \big],  
    \end{align*}
where $\boldsymbol{\mu}_0$ and $\boldsymbol{\Sigma}_0$ parametrize $q_{\boldsymbol{\phi}}(\boldsymbol{z}_i)$, and $\boldsymbol{\mu}_1$ and $\boldsymbol{\Sigma}_1$ parametrize $q_{\boldsymbol{\phi}}(\boldsymbol{z}_i \mid \boldsymbol{A}_i)$. 
The approximated invariant objective, $\mathcal{\tilde{L}}(\boldsymbol{A}_i, y_i, \boldsymbol{c}_i; \boldsymbol{\theta}, \boldsymbol{\phi}) = (1+\lambda)\tilde{\mathcal{L}}_{\text{recon}} - \lambda \tilde{\mathcal{L}}_{\text{marg}} - \tilde{\mathcal{L}}_{\text{prior}}$, is now differentiable w.r.t. $\boldsymbol{\theta}$ and $\boldsymbol{\phi}$. 

\subsection*{\normalfont \textbf{E. Architecture of inv-VAE model}}
The architecture of our inv-VAE is presented in Table \ref{tab:param}. Cross-validation is used to determine the best model for both the ABCD and HCP datasets. The latent dimension size and network depth are chosen to achieve the minimal training loss. We use the Adam optimizer \autocite{kingma2014} to train both inv-VAE and GATE on a GPU device. During model training, the parameters in Table \ref{tab:param} are set to be the same for both inv-VAE and GATE for a fair comparison. The learning rate is set to $5e^{-6}$ in the simulation study and $2e^{-5}$ in applications to the two datasets. For the mini-batches, we choose a batch size of 32, and the batches are sampled uniformly at random at each epoch and repeated for 200 epochs in both simulation study and real applications. 

\textcolor{black}{When the experiment's objective is to adjust for motion-induced artifacts, as demonstrated in Section \ref{sec:experiment_motion}, $90\%$ subjects are included in the training set and $10\%$  subjects are included in the validation set to select optimal model parameters. When the goal is to predict cognition-related traits, as outlined in Section \ref{sec:experiment_trait}, we use 5-fold cross-validation (CV) to designate 80\% of the data as the training and validation data (in which, 90\% are for training the model and 10\% serve as the validation set to choose model parameters) and 20\% as the test set for evaluating model predictions.}

\begin{table}[ht]
  \caption{Model architecture of inv-VAE. }\label{tab:param}
    \vspace{1mm}
    \centering
    \begin{tabular}{|c|c|c|}
    \hline
     & Inference Model (Encoder) & Generative Model (Decoder) \\
     \hline
      & $\boldsymbol{W}_{1}: 68 \times 256$ & $k$-NN: 32\\
     & $\boldsymbol{W}_{2}: 256 \times 68$ & K = 68\\
      & $b_{1}: 256 \times 1$  & $M$ = 2\\
      inv-VAE & $b_{2}: 68 \times 1$ & $R$ = 5\\
       & $\varphi_{1}$: ReLU & $h_1$: Sigmoid\\
        & $\varphi_{2}$: Linear & $h_2$: Sigmoid\\
  \hline
  \multicolumn{3}{p{350pt}}{$K$ is the latent dimension of $\boldsymbol{z}_i$; $M$ is the number of GCN layers; $R$ is the latent dimension of $\boldsymbol{X}(\tilde{\boldsymbol{z}}_i)$; $\boldsymbol{W}_\cdot, b_{\cdot}, \varphi_{\cdot}$ are weights, biases and activation functions in the encoder; $h_{\cdot}$ are the activation functions in the decoder.}
  \end{tabular}
\end{table} 

\subsection*{\normalfont \textbf{F. Correlation between motion score and outlier measure}}

To clarify the specific aspects of image quality captured by our proposed motion measure, we conduct a comparison between the head motion estimate from the eddy tool, the residual misalignment (translation and rotation), and a summary of the outlier measure. In Figure \ref{fig: outlier_eda}, we examine the correlation between the proposed  artifact measures and the number of outlier slices in the HCP dataset. The analysis reveals no significant relationship between outliers and either motion or residual misalignment,  suggesting that the proposed measures capture other distinct artifacts present in the data.

\renewcommand{\thefigure}{A\arabic{figure}}

\setcounter{figure}{0}

\begin{figure}[ht]
\centering
\includegraphics[width=.8\linewidth]{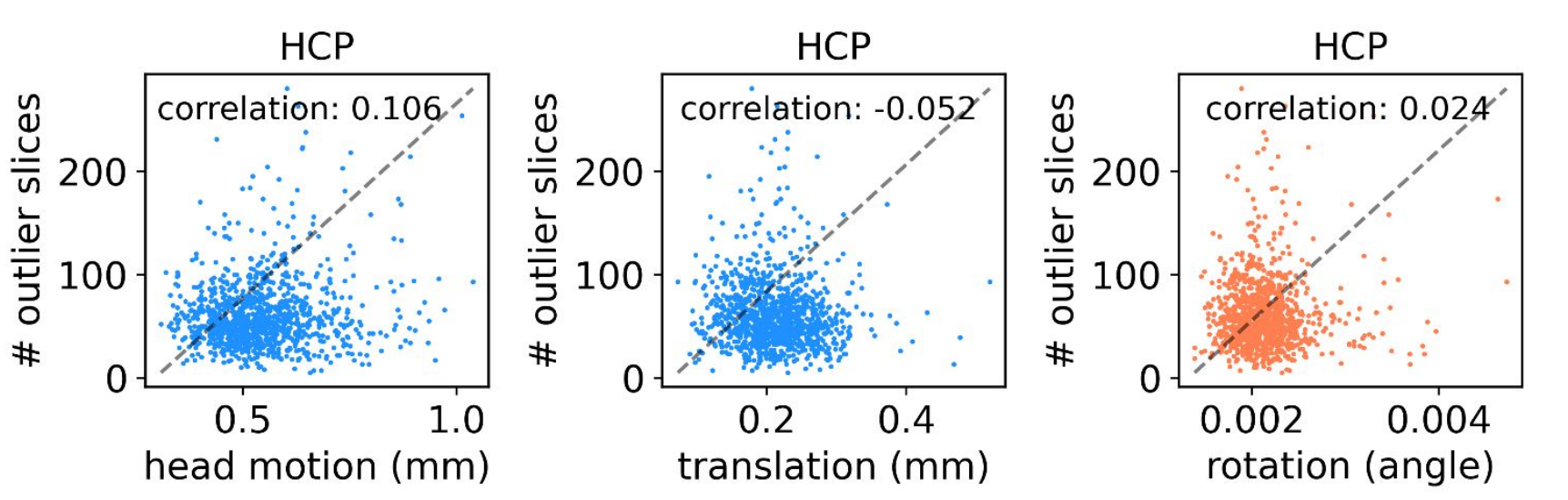}
\caption{Pearson's correlation between head motion, residual misalignment (translation and rotation), and outlier measure in the HCP dataset. \textcolor{black}{Head motion and residual translation is measured in millimeters (mm), while residual rotation is in angles.}} \label{fig: outlier_eda}
\end{figure}


\subsection*{\normalfont \textbf{G. Inv-VAE with residual misalignment adjustment}}

\textcolor{black}{In this section, we apply our proposed inv-VAE to residual misalignment adjustment. Residual misalignment refers to the remaining distortions and imperfections in image registration following FSL eddy correction. Our analyses of the ABCD and HCP datasets indicate that, although residual misalignment exhibits a low correlation with head motion obtained from FSL eddy (Figure \ref{fig: motion_residual_corr}), it can still impact the constructed brain networks. Therefore, addressing this artifact is crucial as it enables us to correct for sources of artifacts that are not correlated with head motion.} 

\textcolor{black}{To measure the residual misalignment, we rely on $b=0$ images (images acquired without
diffusion weighting). In the ABCD data, each individual has 7 $b=0$ frames evenly distributed throughout the diffusion scan, while each HCP subject has 18 frames with $b=0$. We use the FSL FLIRT tool to align other $b=0$ images to the first $b=0$ image with 6 degrees of freedom (DOF = 6, rigid body). Our analysis employs FSL version 5.0.9 for both the eddy and FLIRT tools. To quantify residual misalignment, we use the total amplitudes of rotation and translation. For each individual, we calculate the average translation and rotation displacements across all frames, providing a measure of residual misalignment. The detailed process for extracting these representations of residual translation and rotation is described in Appendix Section I.} 

\begin{figure}
\centering
\includegraphics[width=.85\linewidth]{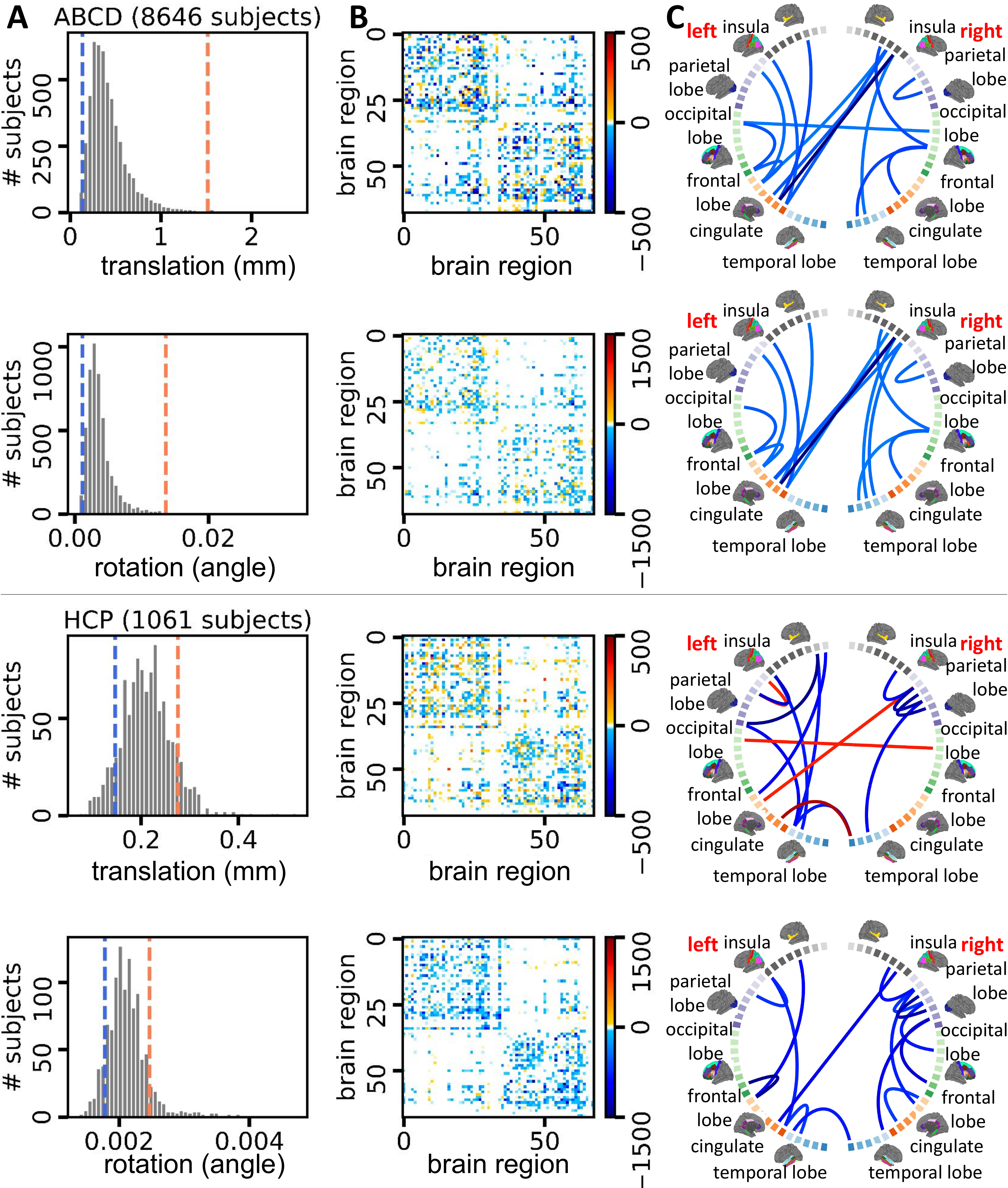}
\caption{(\textbf{A}) \textcolor{black}{Distributions of residual translation in millimeters (mm) and residual rotation in angles for the ABCD and HCP datasets.} In ABCD data, the 5th and 95-th percentiles are marked by the blue and orange lines. In HCP data, the blue and orange lines indicate the 10-th and 90-th percentiles. Subjects whose residual translation (rotation) fall below the blue line are assigned to the small translation (rotation) group, and those falling above the orange line are in the large translation (rotation) group. (\textbf{B}) Differences of mean networks between the large and small (large $-$ small) translation (rotation) groups. (\textbf{C}) Plotting column (\textbf{B}) in circular plots. The color scale is the same as that in (\textbf{B}), but we only show 15 edges with the largest fiber count differences for display.} \label{fig: eddy_eda}
\end{figure}

\begin{figure}[ht]
\centering
\includegraphics[width=\linewidth]{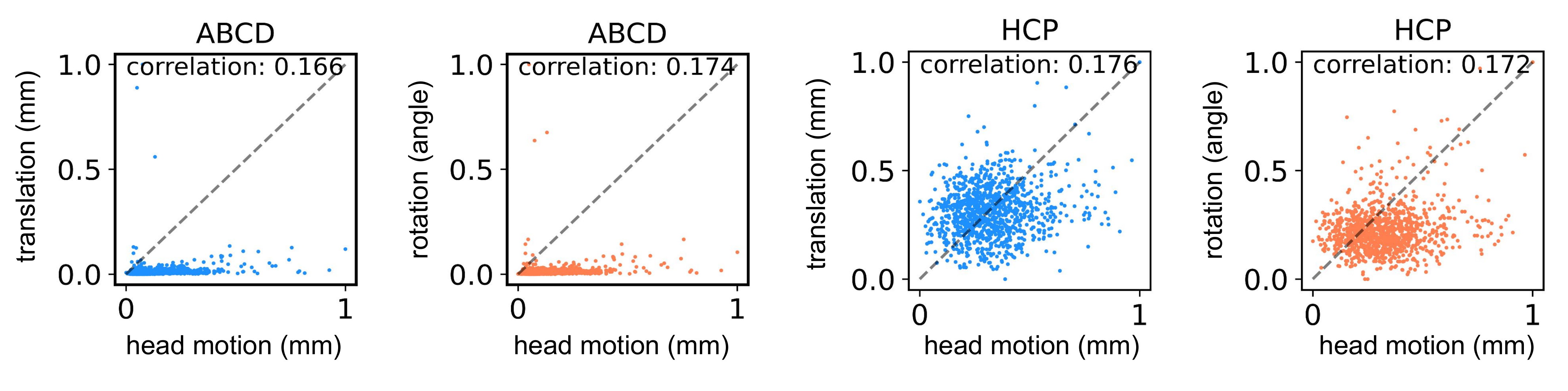}
\caption{Pearson's correlation between head motion and the residual translation and rotation in each dataset. Head motion is quantified in millimeters (mm) by computing the average displacement of each intracerebral voxel, then calculating the square root of the average squared displacements. Residual translation is measured in millimeters, while residual rotation is in angles.} \label{fig: motion_residual_corr}
\end{figure}

\textcolor{black}{The residual translation and rotation in the ABCD and HCP datasets demonstrate a relatively low correlation with the head motion (see Figure \ref{fig: motion_residual_corr}), motivating us to check further the relationship between the residual misalignment and brain structural network. The results are displayed in Figure \ref{fig: eddy_eda} (\textbf{A})-(\textbf{C}). In the column \textbf{A} of Figure \ref{fig: eddy_eda}, we display the histograms of residual translation and rotation for ABCD and HCP subjects, with the blue and orange lines representing the 10th and 90th percentiles. Subjects with the residual translation (rotation) below the blue line are classified into the small translation (rotation) group, while those above the orange line belong to the large translation (rotation) group. In column (\textbf{B}), we show the mean network differences between the large and small translation (rotation) groups (large $-$ small). Column (\textbf{C}) shows circular plots representing the data in panel (\textbf{C}), limited to the 15 edges showing the largest fiber count differences.}

\begin{figure}
\centering
\includegraphics[width=.8\linewidth]{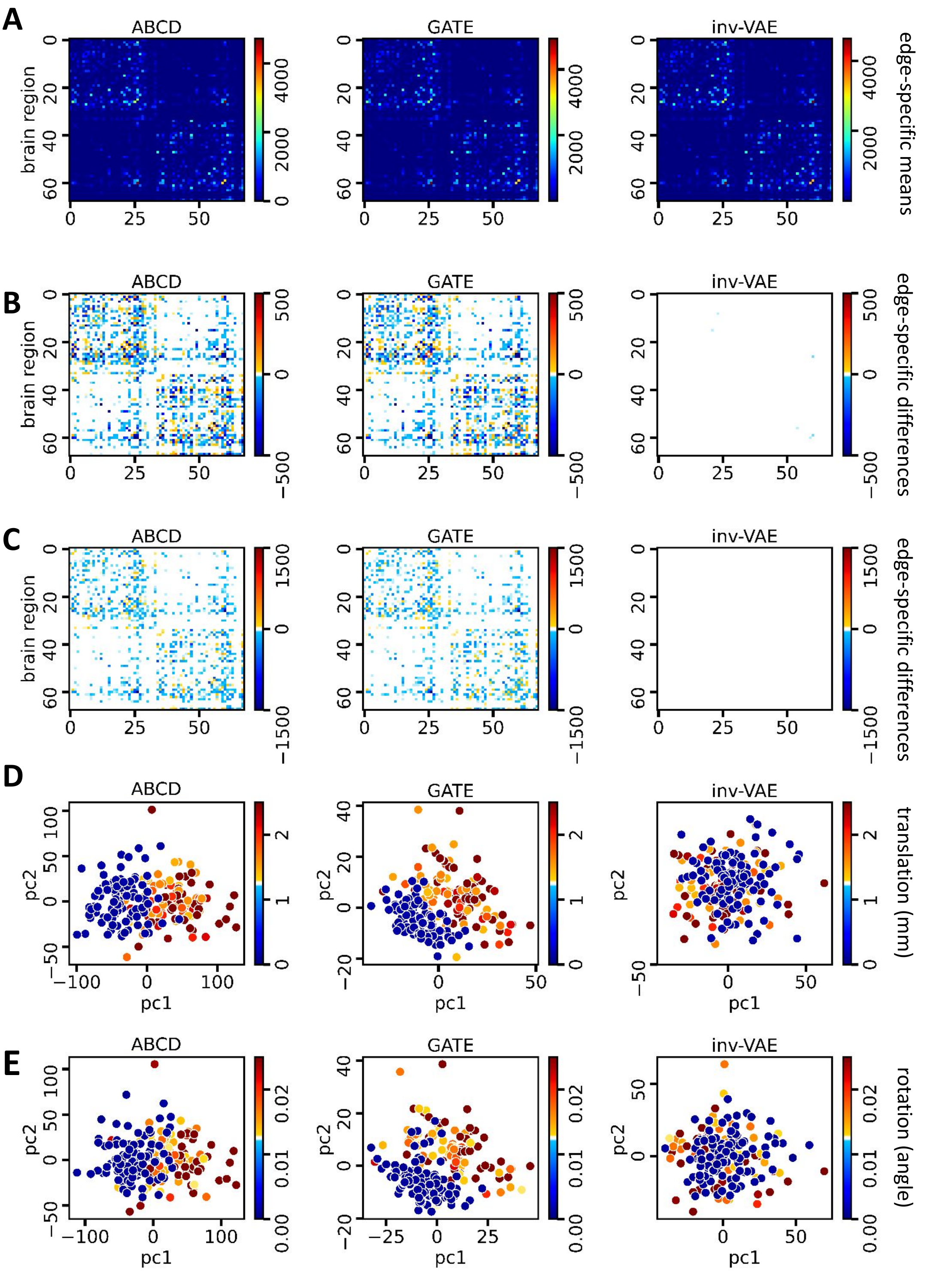}
\caption{(\textbf{A}) From left to right: observed edge-specific means of ABCD brain networks, reconstructed brain networks by GATE, and residual-adjusted brain networks by inv-VAE, color-coded by fiber count. (\textbf{B}) and (\textbf{C}):  edge-specific differences (large group subtract small group) for residual translation and rotation respectively, colored by differences in fiber count. Ordering follows that in (\textbf{A}). (\textbf{D}) and (\textbf{E}): from left to right - the first two PC scores of ABCD brain networks, latent embeddings learned by GATE, and invariant embeddings from inv-VAE for observations in the large and small translation (rotation) groups, colored by the amount of residual translation (\textbf{D}) and rotation (\textbf{E}) respectively. \textcolor{black}{Residual translation is in millimeters (mm) and residual rotation is in angles.}} \label{fig:abcd_resid_adjust}
\end{figure}

\begin{figure}
\centering
\includegraphics[width=.8\linewidth]{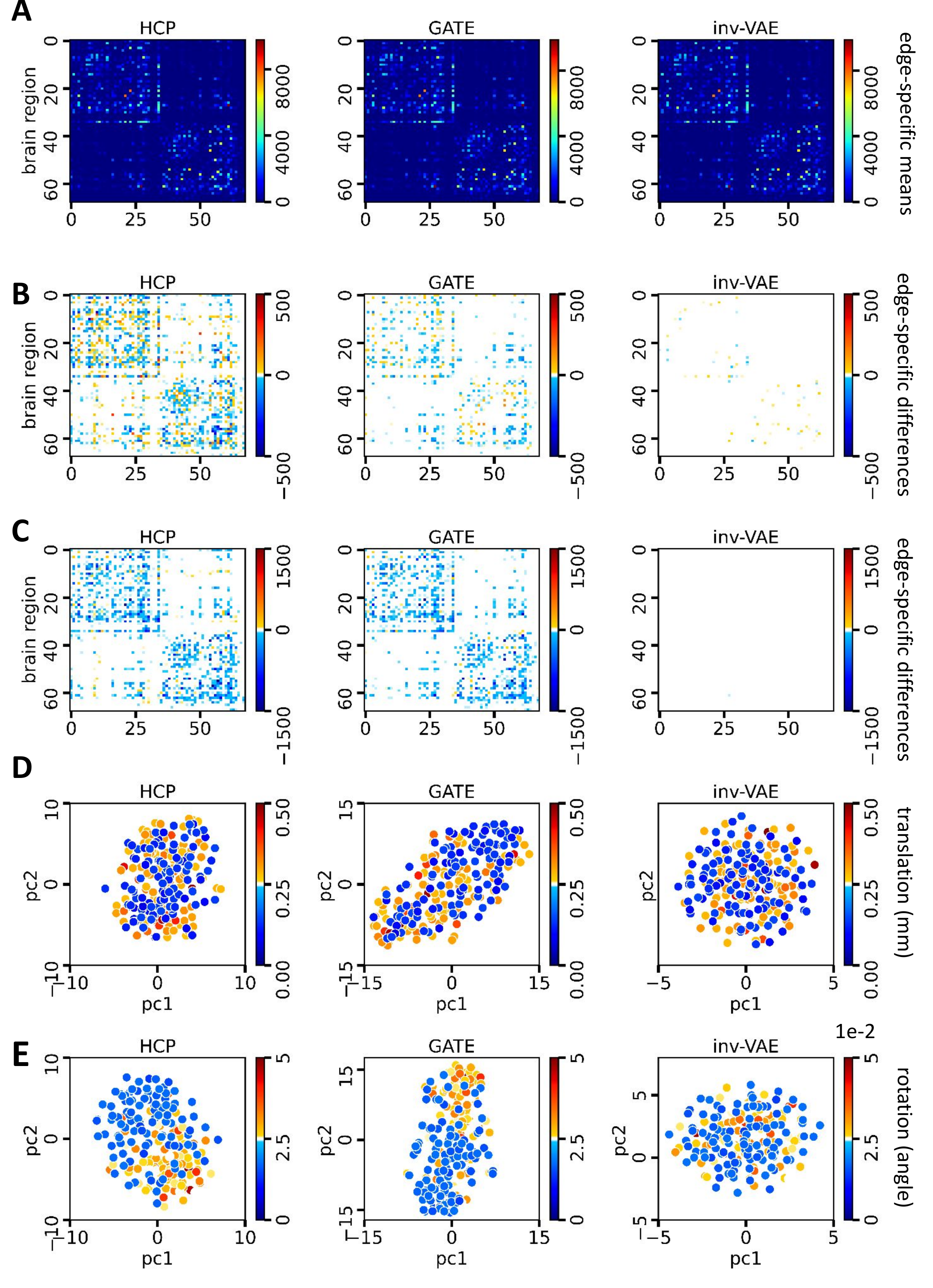}
\caption{(\textbf{A}) From left to right: observed edge-specific means of HCP brain networks, reconstructed brain networks by GATE, and residual-adjusted brain networks by inv-VAE, color-coded by fiber count. (\textbf{B}) and (\textbf{C}): edge-specific differences (large group subtract small group) for residual translation and rotation respectively, colored by differences in fiber count. Ordering follows that in (\textbf{A}). (\textbf{D}) and (\textbf{E}): from left to right - the first two PC scores of HCP brain networks, latent embeddings learned by GATE, and invariant embeddings from inv-VAE for observations in the large and small translation (rotation) groups, colored by the amount of residual translation (\textbf{D}) and rotation (\textbf{E}) respectively. \textcolor{black}{Residual translation is in millimeters (mm) and residual rotation is in angles.}} \label{fig:hcp_resid_adjust}
\end{figure}

\textcolor{black}{We further conduct a similar analysis parallel with Section 5.1 ``Understanding and Removing Motion Artifacts'' to demonstrate the impact of residual misalignment and mitigate its effects on both the ABCD and HCP datasets. Firstly, we average the residual-affected brain network $\boldsymbol{A}_i$ across all individuals in the study, obtaining the residual-affected edge-specific means for both datasets. The first columns of Figure \ref{fig:abcd_resid_adjust} and \ref{fig:hcp_resid_adjust} (\textbf{A}) illustrate these residual-affected edge-specific means in both datasets. 
The first columns of Figure \ref{fig:abcd_resid_adjust} and \ref{fig:hcp_resid_adjust} (\textbf{B}, \textbf{C}) display the differences in residual-affected edges in ABCD and HCP data, determined by subtracting the means of the small translation (rotation) group from those of the large translation (rotation) group. Additionally, we visualize the first two principal components (PCs) of brain networks in the first columns of Figure \ref{fig:abcd_resid_adjust} and \ref{fig:hcp_resid_adjust} (\textbf{D}, \textbf{E}), with each point representing an individual colored according to their corresponding residual translation (rotation) displacement.}

\textcolor{black}{We then apply GATE \autocite{liu2019} to both the ABCD and HCP datasets. The second columns of Figure \ref{fig:abcd_resid_adjust} and \ref{fig:hcp_resid_adjust} (\textbf{A})  illustrate the reconstructed edge-specific means averaged across all reconstructed networks $\hat{\boldsymbol{A}_i}$ for both datasets. The second columns of Figure \ref{fig:abcd_resid_adjust} and \ref{fig:hcp_resid_adjust} (\textbf{B}, \textbf{C}) show reconstructed edge-specific differences between the large and small translation (rotation) groups. The first two PCs of the learned GATE latents from the large and small translation (rotation) groups are displayed in the second columns of Figure \ref{fig:abcd_resid_adjust} and \ref{fig:hcp_resid_adjust}  (\textbf{D}, \textbf{E}) for residual translation (rotation). }

\begin{figure}[t]
\centering
\includegraphics[width=\linewidth]{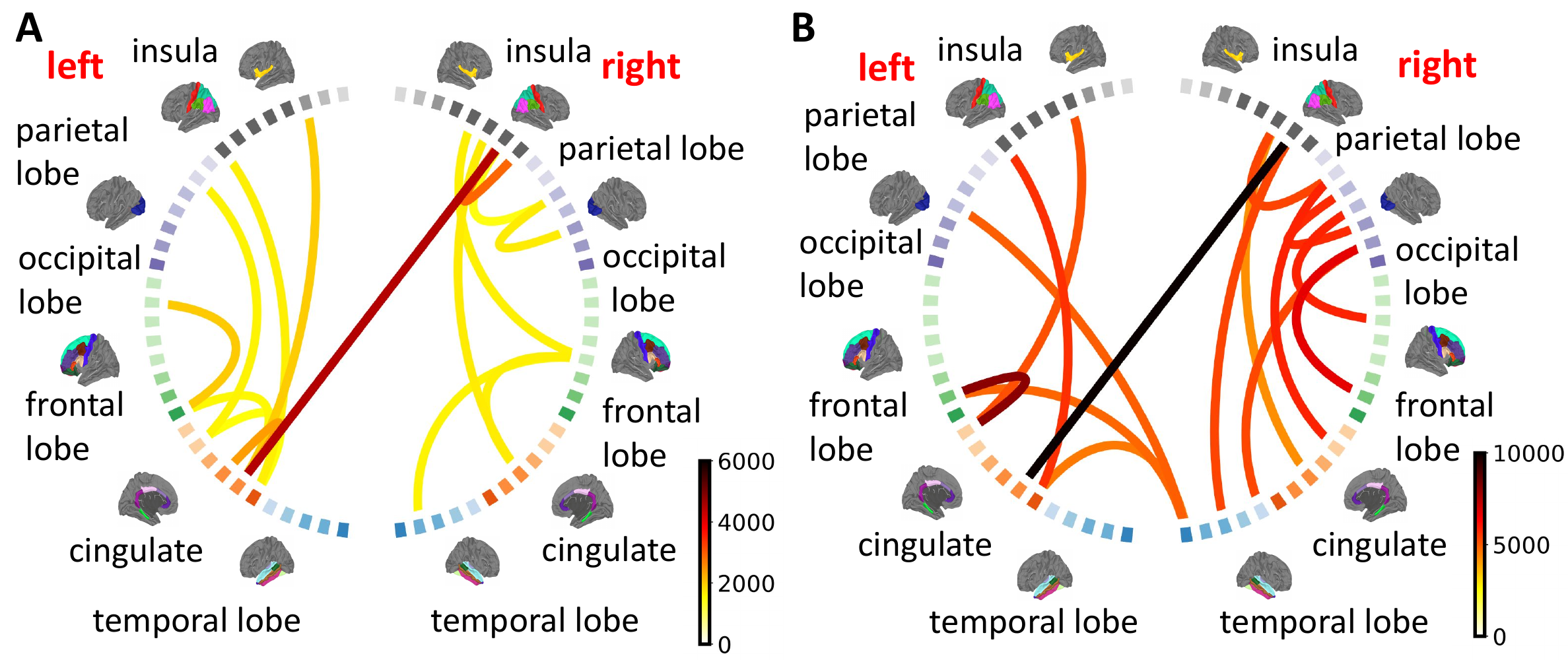}
\caption{Structural connectivity differences between residual-adjusted and residual-affected ABCD (\textbf{A}) and HCP (\textbf{B}) brain networks. Each line connecting a pair of brain regions is colored by the corresponding fiber count difference after adjusting for residual misalignment. We only show the 15 mostly changed brain connections.}\label{fig: resid_circ_plots}
\end{figure}

\begin{figure}[t]
\centering
\includegraphics[width=.9\linewidth]{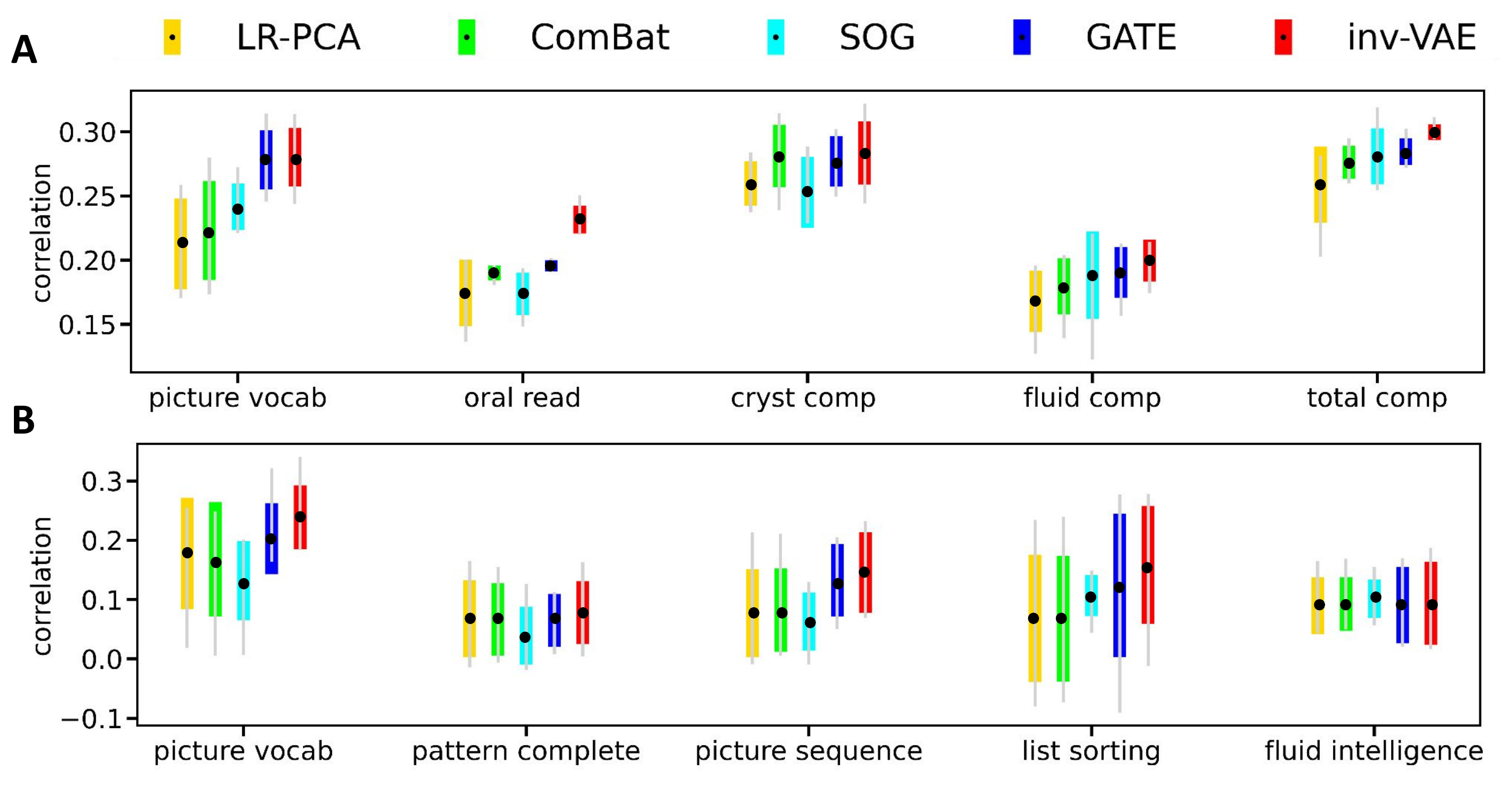}
\caption{A comparison of all methods on trait prediction in the ABCD (\textbf{A}) and HCP (\textbf{B}) data. The error bars show the minimum, maximum, mean, and standard deviation of correlation coefficients from a 5-fold CV. We pre-process the target trait $y_i$ by regressing out residual misalignment. For each fold, we train inv-VAE on the training data. For the test data, we obtain trait predictions after setting $c_i$ to 0 to remove residual misalignment.}  \label{fig:resid_pred}
\end{figure}

\begin{figure}[t]
\centering
\includegraphics[width=\linewidth]{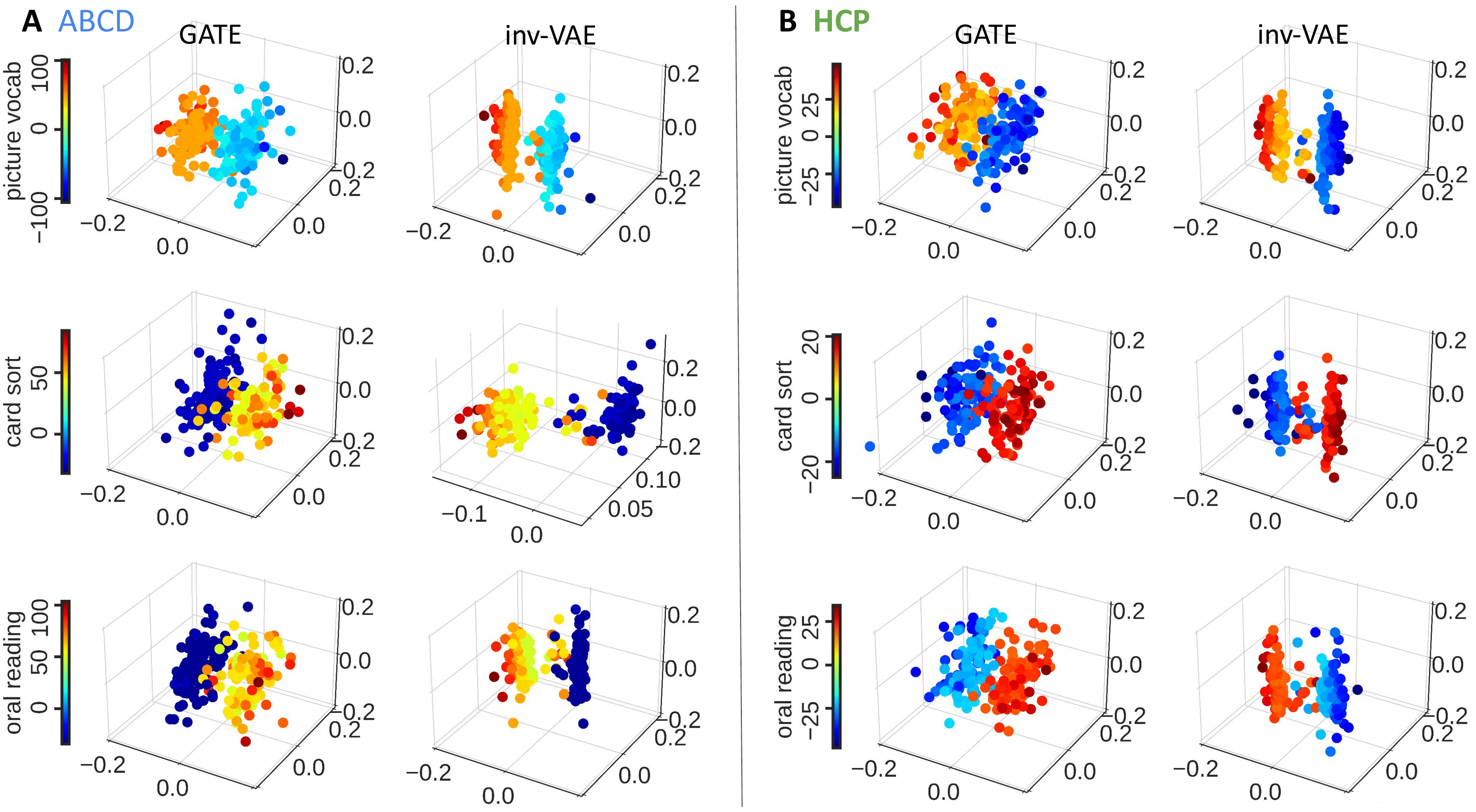}
\caption{Relationships between learned embeddings and cognition-related traits in ABCD (\textbf{A}) and HCP (\textbf{B}) datasets. {We pre-process the target trait $y_i$ by regressing out residual misalignment. Both inv-VAE and GATE are trained with brain networks of all individuals in each dataset to obtain the latent $\boldsymbol{z}_i$'s for the $y_i$'s. For each trait, latent features from 200 subjects are selected, with the first group of 100 having the lowest trait scores and the second group of 100 having the highest scores. The first three PCs of the posterior means of $\boldsymbol{z}_i$ colored with their corresponding trait scores are displayed for each trait $y_i$.} Latent embeddings produced by GATE and residual-invariant embeddings from inv-VAE are compared.}\label{fig:resid_pca}
\end{figure}

\textcolor{black}{To address artifacts due to residual misalignment, we train inv-VAE separately using the ABCD and HCP data. We generate residual-adjusted brain networks by setting $\boldsymbol{c}_i = (c_{i1}, c_{i2})$ to 0, where $c_{i1}$ and $c_{i2}$ represent residual translational and rotational displacements, to eliminate residual artifacts. Averaging across all residual-adjusted networks for each dataset yields the residual-adjusted edge-specific means shown in the third columns of Figure \ref{fig:abcd_resid_adjust} and \ref{fig:hcp_resid_adjust} (\textbf{A}). The third columns of Figure \ref{fig:abcd_resid_adjust} and \ref{fig:hcp_resid_adjust} (\textbf{B}, \textbf{C}) illustrate the residual-adjusted edge-specific differences obtained by subtracting the residual-adjusted edge-specific means of the large and small translation (rotation) groups. The third columns of Figure \ref{fig:abcd_resid_adjust} and \ref{fig:hcp_resid_adjust} (\textbf{D}, \textbf{E}) depict the first two PCs of invariant representations $\boldsymbol{z}_i$ for both datasets. The gap between large and small translation (rotation) groups is reduced after residual adjustment compared to the projections of the observed brain networks.}

\textcolor{black}{We further investigate how residual artifacts influence our understanding of structural brain connectivity. For both ABCD and HCP datasets, we first compute the residual-affected edge-specific means by averaging $\boldsymbol{A}_i$'s across all subjects in the dataset. After training inv-VAE using all $\boldsymbol{A}_i$'s, we generate residual-adjusted networks $\boldsymbol{A}^*_i$'s from the invariant embeddings. We then average across all $\boldsymbol{A}^*_i$'s to obtain the residual-adjusted edge-specific means. Figure \ref{fig: resid_circ_plots} (\textbf{A}, \textbf{B}) illustrate the differences between the residual-adjusted and residual-affected edge-specific means.}

\textcolor{black}{Parallel to Section \ref{sec:experiment_trait}, we explore whether residual-invariant representations can improve our understanding of the relationship between brain connectomes and cognition-related traits. To prevent mistakenly attributing connectome differences to residual misalignment rather than underlying traits, we predict the residuals of the trait $y_i$ after regressing out the residual artifacts, $\textbf{c}_i = (c_{i1}, c_{i2})$, where $c_{i1}$ and $c_{i2}$ are the mean translational and rotational displacements across all frames. Figure \ref{fig:resid_pred} compares inv-VAE's trait-prediction quality to other baselines described in Section \ref{sec:experiment_trait}. The results demonstrate that inv-VAE outperforms LR-PCA, ComBat, and SOG across all studied traits in both datasets, while either surpassing or matching GATE's performance for most selected traits.}

\textcolor{black}{We also assess predictive performance by visualizing associations between latent representations of structural connectomes and cognitive traits. For both ABCD and HCP studies, we select specific traits and use inv-VAE and GATE to produce latent features $\boldsymbol{z}_i$ for each subject. We then choose 200 subjects (100 with the lowest scores, 100 with the highest) for each trait and visualize the first three PCs of the posterior means of their $\boldsymbol{z}_i$, colored by trait scores (Figure \ref{fig:resid_pca}). These plots show separation between high and low trait groups, indicating structural connectome differences related to cognitive abilities. Notably, inv-VAE's residual-invariant $\boldsymbol{z}_i$'s show clearer separation than GATE's residual-affected $\boldsymbol{z}_i$'s, especially for the oral reading recognition test. This suggests residual-invariant structural connectomes have stronger associations with cognitive traits compared to approaches without residual misalignment adjustment.}


\subsection*{\normalfont \textbf{H. Extreme motion and breakdown point}}
\textcolor{black}{Given that HCP and ABCD data undergo multiple curation steps to exclude instances of very high motion, the threshold  at which our proposed method might encounter limitations and fail to correct for extreme subject motion remains unclear. To address this, we conduct tests on our motion-invariant method using simulated data with varying levels of nuisance variables, evaluating the method's robustness. }

\begin{figure}[ht]
\centering
\includegraphics[width=\linewidth]{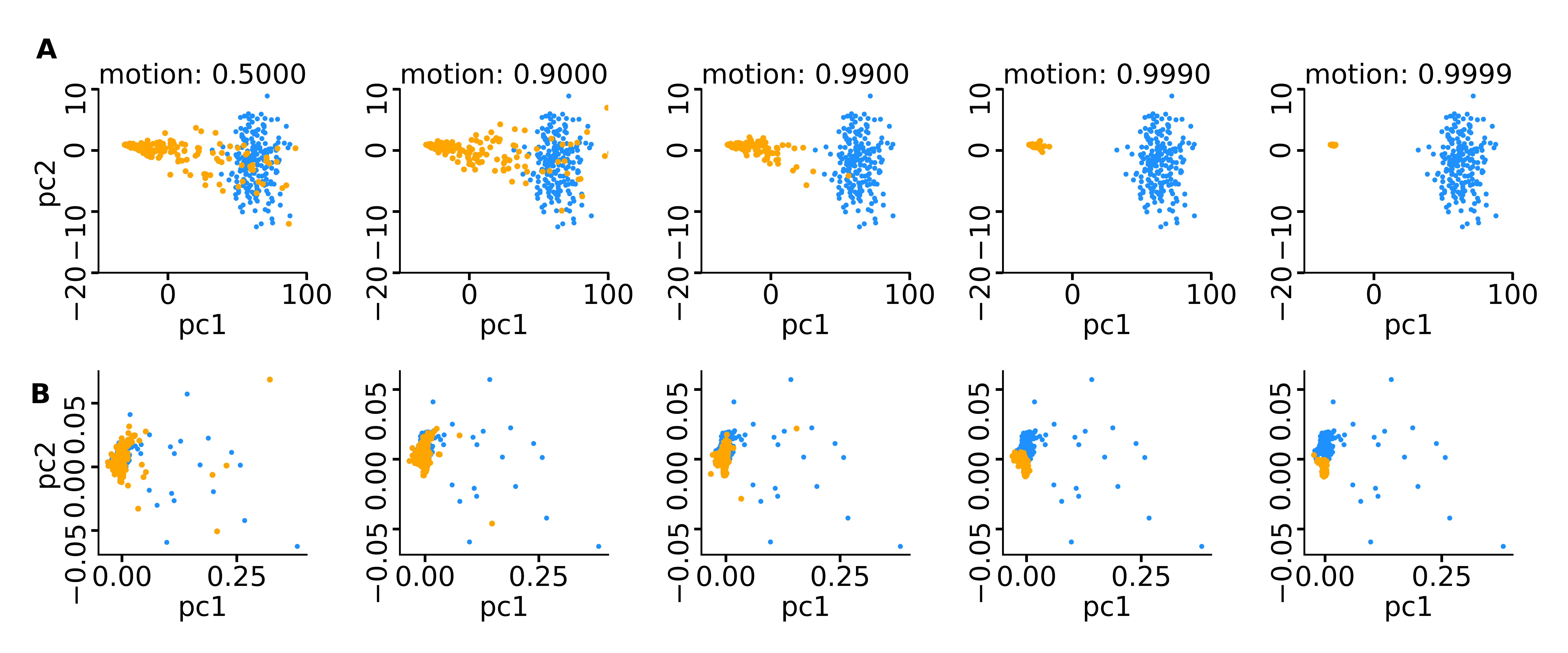}
\caption{Analysis of the robustness of the proposed method under extreme motion and motion invariance test. (\textbf{A}) The first two PCs of the nuisance-free (blue) and the nuisance-affected (orange) networks under various levels of nuisances. (\textbf{B}) The first two PCs of the learned invariant latents for the nuisance-free (blue) and the nuisance-affected (orange) networks. } \label{fig: extreme_motion}
\end{figure}

\textcolor{black}{We expand the simulation study outlined in Section 4 with varying motion levels. In particular, we simulate networks $\{\boldsymbol{A}_i\}_{i=1}^{n}$, where $\boldsymbol{A}_i \in \mathbb{R}^{V\times V}$ with $V = 68$ nodes and $n = 200$. We then  introduce a nuisance variable, $\{s_i\}_{i=1}^n$, by sampling its elements from $\mathcal{N}(\mu_j, \mu_j)$ with $\mu_j \in \{0.5, 0.1, 0.01, 0.001, 0.0001\}$. {To keep the notation consistent, we use $c_i$ to denote the simulated motion, and set $c_i = 1 - s_i$. Motion-affected networks $\tilde{\boldsymbol{A}_i}$ are generated by propagating $c_i$ across all edges in $\boldsymbol{A}_i$ using the expression:
\begin{align}
\tilde{\boldsymbol{A}_i} = ((1 - c_i) \boldsymbol{A}_i)^\top ((1 - c_i) \boldsymbol{A}_i). \label{eq:sim_bkp}
\end{align}
Intuitively, large $c_i$'s introduce large artifacts to the simulated networks, as the multiplication of $1 - c_i$ with $\boldsymbol{A}_i$ in Equation (\ref{eq:sim_bkp}) results in a $\tilde{\boldsymbol{A}_i}$ with greatly reduced edge values; small $c_i$'s generate smaller artifacts as the multiplication operation in Equation (\ref{eq:sim_bkp}) changes the edges of $\boldsymbol{A}_i$ to a lesser extent. Figure \ref{fig: extreme_motion} (\textbf{A}) displays the first two principal components of the simulated brain networks colored by the motion group (motion-free is colored blue and motion-affected is colored orange), illustrating that as the simulated motion $c_i$ increases, the distance between the two groups of networks becomes larger, making it more challenging for our proposed method to mitigate distortions from such artifacts.} }

\textcolor{black}{We then train inv-VAE on the simulated motion-free and motion-affected networks to learn the parameters $\boldsymbol{\phi}$ and $\boldsymbol{\theta}$ to learn the motion-invariant latent representations (see Figure \ref{fig: extreme_motion} (\textbf{B})). As defined earlier, $c_i \approx 0$ corresponds to small motion artifacts, whereas large $c_i$'s represent large motion artifacts. If our method cannot accurately recover the motion-free $z_i$, the PCs of the invariant $z_i$'s from the motion-free and motion-affected group will be separated. In Figure \ref{fig: extreme_motion} (\textbf{B}), we visually display the invariant $z_i$'s of the two groups. Notably, when $c_i$ is small, our proposed method performs well in recovering the motion-free latents. However, when $c_i$ is large, our proposed method starts to break down, indicated by the clear separation between the two groups of latent representations.} 

\textcolor{black}{We have conducted a simple motion-invariance test using both ABCD and HCP data. In Figures 4 and 5  panels (\textbf{D}-\textbf{E}) in the manuscript, we present the first two principal component (PC) scores of observed brain networks, latent embeddings acquired through GATE, and invariant embeddings obtained from inv-VAE for brain networks in both large and small-motion groups. Notably, we observe that both the PCs of the observed brain networks and GATE latents are correlated with the subject motion $c_i$. In contrast, the invariant $z_i$ demonstrate independence from $c_i$.}

\textcolor{black}{Additionally, we conduct a simulation study to see if the same subject, imaged with varying amounts of motion would have the same $z_i$ vector. For a given $A_i$, we can simulate various motion-affected networks $\tilde{A}_i$ with different motion parameters $c_i$. To check whether identical $z_i$ can be obtained from these $\tilde{A}_i$'s, we generate $\tilde{A}_i$ with a distinct $c_i = 1 - s_i$ and examine whether the embeddings match those obtained when $c_i = 0$. To test whether identical networks yield the same $z_i$ under different motion levels, we use the symmetric Kullback-Leibler (KL) divergence to quantify the distance between the learned $z_i$'s. This choice is motivated by the fact that each $z_i$ follows a multivariate normal distribution characterized by means and variances learned by our inference model:
\begin{align*}
KL = \frac{1}{2}(D_{KL}(P || Q) + D_{KL}(Q || P)),
\end{align*}
where $P$ and $Q$ are the distributions of the motion-free $z_i$ and motion-invariant $z_i$, respectively. If our method (inv-VAE) cannot accurately recover the same $z_i$, the network will have a large KL value.}

\begin{figure}
\centering
\includegraphics[width=\linewidth]{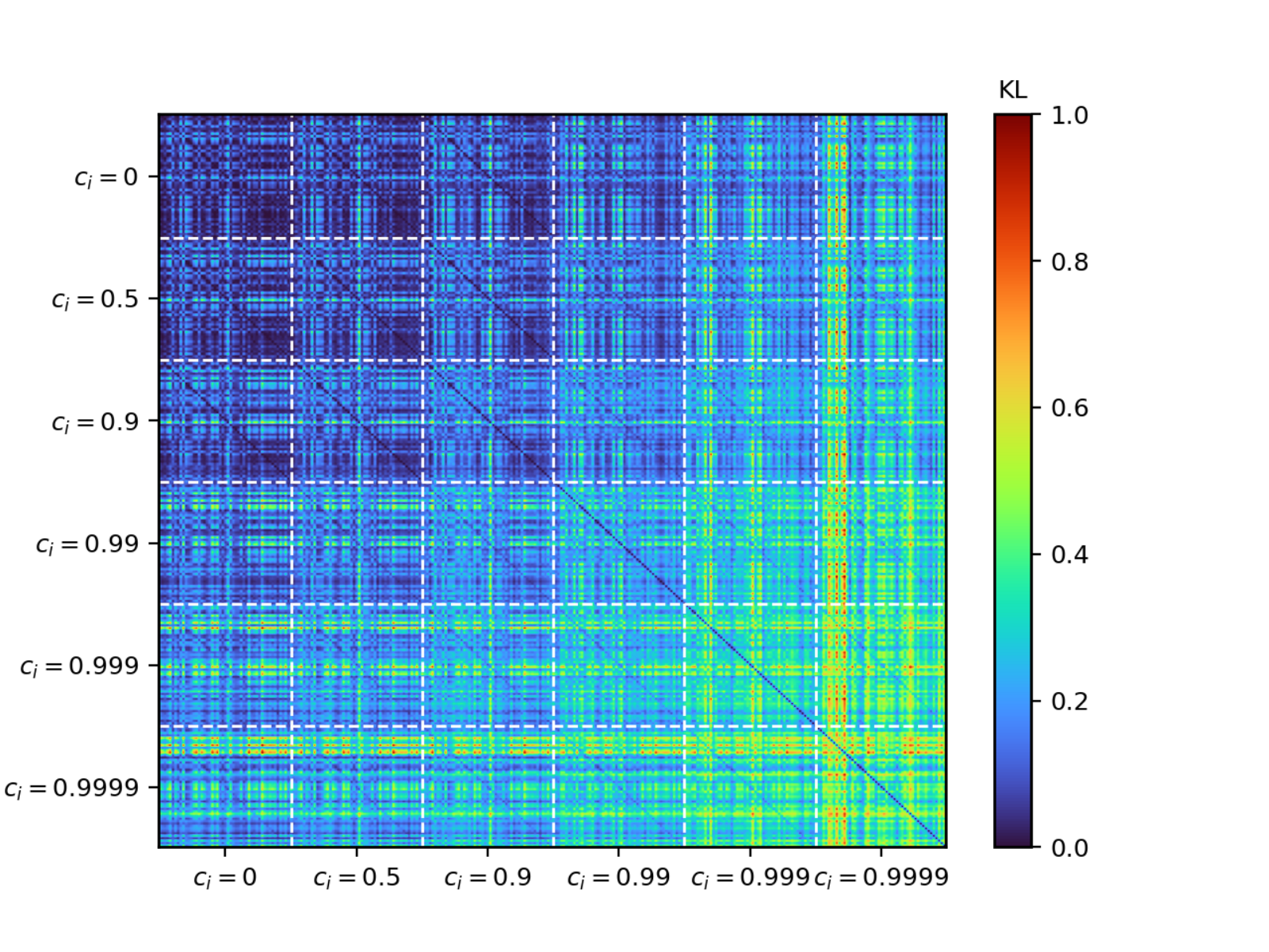}
\caption{We construct a distance matrix to measure the KL divergence between each pair of $z_i$'s within and between the corresponding motion groups. Here, $c_i = 0$ denotes the simulated motion-free networks. We simulate motion-affected networks with 5 levels of motion $c_i = 1 - s_i \in \{0.5, 0.9, 0.99, 0.999, 0.9999\}$. Each motion group comprises 200 networks. Each matrix entry represents the KL divergence between a pair of networks, either within the same or different motion groups. The colors in the matrix reflect the normalized KL divergence.}  \label{fig: kl_distance}
\end{figure}

\textcolor{black}{In our simulation, we set $c_i = 0$ to simulate motion-free networks, and simulate motion-affected networks with 5 levels of motion $c_i = 1 - s_i \in \{0.5, 0.9, 0.99, 0.999, 0.9999\}$. Each motion group comprises 200 networks. We construct a distance matrix to measure the KL divergence between each pair of $z_i$'s within and between the corresponding motion groups. In Figure \ref{fig: kl_distance}, each entry of the distance matrix represents the KL divergence between a pair of networks, either within the same or different motion groups. For visualization purposes, we normalize the KL divergence across network pairs within each sub-matrix in Figure \ref{fig: kl_distance}. Examining the first row (column) of the distance matrix allows us to assess the motion invariance of $z_i$'s across various motion conditions—ranging from no motion ($c_i = 0$) to increased levels of motion ($c_i \in \{0.5, 0.9, 0.99, 0.999, 0.9999\}$). Remarkably, when motion is not large, e.g., $c_i \in \{0.5, 0.9\}$, our proposed method successfully recovers the same $z_i$ vector, evident from the distinct dark diagonal line in the sub-matrices in the first column (row) of Figure \ref{fig: kl_distance} (indicating low KL divergence, approximately 0). However, as motion levels ($c_i$) increase, our proposed method exhibits limitations, illustrated by the increased KL values between the motion-free $z_i$ and the learned motion-invariant $z_i$.}


\subsection*{I. Representation of residual misregistration}

For residual translation, it is represented as a 3D vector and we compute its amplitude using an $L_2$ norm. For residual rotation, we utilize the method introduced in \autocite{zhang2019nonparametric}. More specifically, let ${ {\bf I}_3}$ denote the identity element of rotation matrices $SO(3)$. The tangent space at ${ {\bf I}_3}$, $T_{ {\bf I}_3}{(SO(3))}$ forms a Lie algebra, which is usually denoted as $\mathfrak{so}(3)$. The exponential map, $\exp: \mathfrak{so}(3) \rightarrow SO(3)$, provides a mapping from the tangent space $T_{ {\bf I}_3}({SO(3)})$ to $SO(3)$. The inverse of the exponential map is called the $\log$ map. $\mathfrak{so}(3)$ is a set of $3\times 3$ skew-symmetric matrices. We use the following notation to denote any matrix ${{{\bf P}_v} }\in\mathfrak{so}(3) $:
\begin{equation} \nonumber
{{\bf P}_{v}} = 
 \begin{pmatrix}
  0 & -v_1 & v_2  \\
  v_1 & 0& -v_3  \\
  -v_2 & v_3 & 0
 \end{pmatrix}
 \end{equation}
 where ${v}=[v_1,v_2,v_3] \in \mathbb{R}^3$.  The exponential map for $\mathfrak{so}(3)$ is given by Rodrigues' formula, 
 \[ \exp({ \bf P_v}) =
  \begin{cases}
    { \bf I}_3,      & \quad \alpha = 0\\
    {\bf I}_3 + \frac{\sin(\alpha)}{\alpha} { {\bf P}_v} + \frac{1-cos(\alpha)} {\alpha^2}  { {\bf P}_v} ^2,  & \alpha \in (0,\pi) , \\
  \end{cases}
\]
where $\alpha = \sqrt{{1\over 2}  tr ({ {\bf P}_v}^T{{\bf P}_v } )} = \| v\| \in [0,\pi)$. The $\log$ map for a matrix ${ \bf X} \in SO(3)$ is a matrix in $\mathfrak{so}(3)$, given by 
 \[ \log({ \bf X}) =
  \begin{cases}
    {\bf 0}, & \quad \alpha = 0\\
    {\alpha \over 2 \sin(\alpha)}({ \bf X}-{\bf X}^T,) & |\alpha| \in (0,\pi) , \\
  \end{cases}
\]
where $\alpha$ satisfies $tr({ \bf X})=2\cos(\alpha) + 1$. Define a mapping $\Phi$ to embed an element in $\mathfrak{so}(3)$ to $\mathbb{R}^3$, $\Phi: \mathfrak{so}(3) \rightarrow \Re^3$, $\phi({ \bf A}_v) = { v}$. The vector $v$ is used in our paper to denote the residual rotation. 

This is essentially an axis-angle representation of a rotation, which converts a rotation matrix into a compact, vector form that is easier to work with in many applications. To link this mapping to Euler angles, one would need to perform additional transformations, because Euler angles are a different way of representing 3D rotations. Euler angles specify three sequential rotations around the axes of a coordinate system, typically referred to as roll, pitch, and yaw.

\begin{figure}[ht]
\centering
\includegraphics[width=\linewidth]{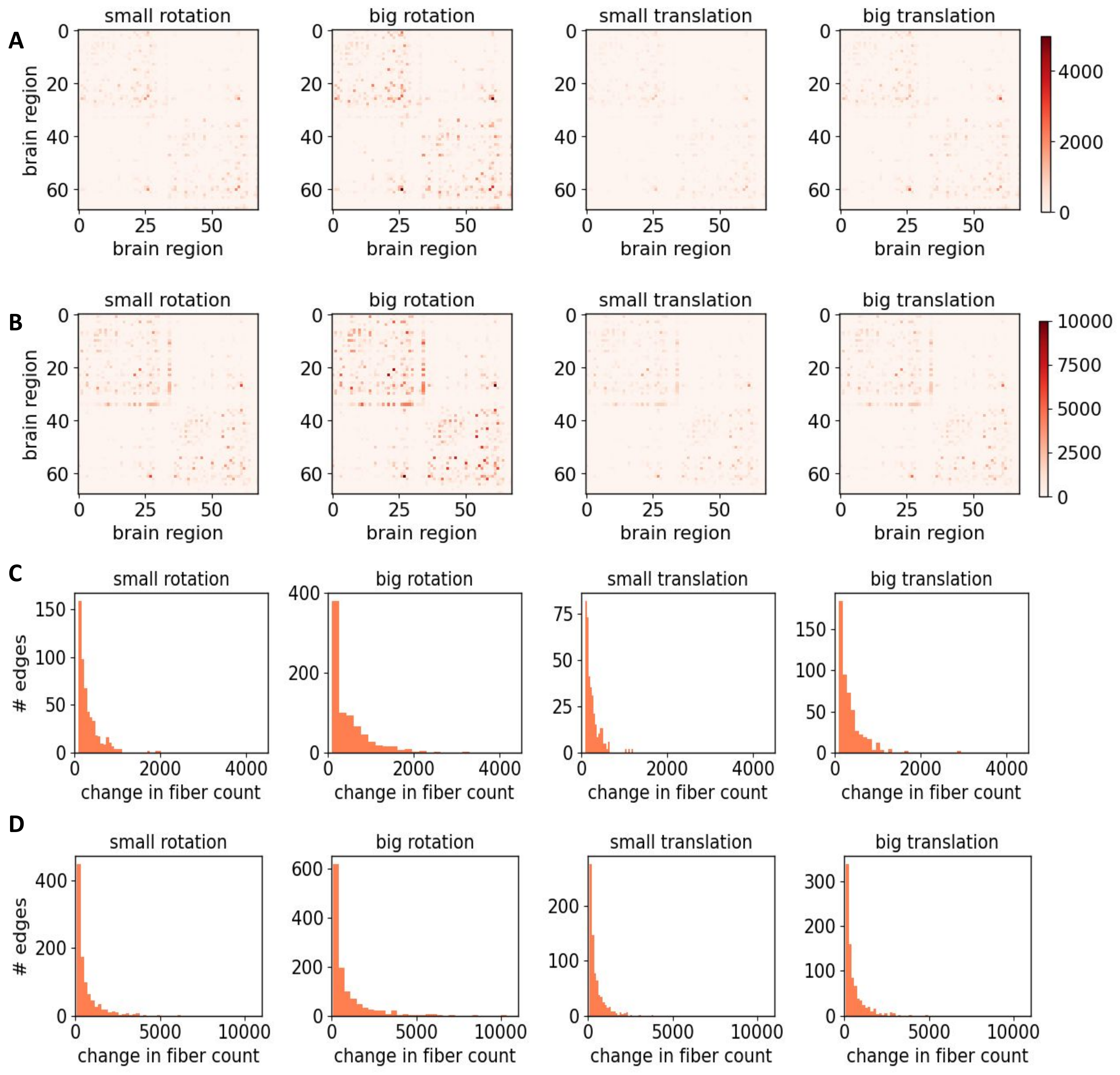}
\caption{When setting $c_i = 0$ and applying the generative model, we observe small differences in the connectomes for the population with low residual artifacts, whereas large differences are observed in the connectomes of the populations with high residual artifacts. Edge-specific differences between the residual-adjusted networks and the original ABCD (\textbf{A}) and HCP (\textbf{B}) networks are averaged across all subjects in the small and large residual translation (rotation) groups. Colors represent the change in fiber counts after residual misalignment correction. {The histograms depicting the scale of correction in fiber counts in the ABCD (\textbf{C}) and HCP (\textbf{D}) data are also presented. To improve clarity in visualization, only brain connections with a change in fiber counts exceeding 100 are displayed.}} \label{fig: absolute_motion_scale}
\end{figure}

\end{doublespace}

\end{document}